\documentclass[twocolumn,showpacs,prl,superscriptaddress]{revtex4-1}%
\usepackage{amsfonts}
\usepackage{amsmath}
\usepackage{amssymb}
\usepackage{graphicx}%
\providecommand{\U}[1]{\protect\rule{.1in}{.1in}}

\begin{document}
\title{Quantum theory of nuclear spin dynamics in diamond nitrogen-vacancy center}
\author{Ping Wang}
\affiliation{Hefei National Laboratory for Physics Sciences at Microscale and Department of
Modern Physics, University of Science and Technology of China, Hefei, Anhui
230026, China}
\affiliation{Beijing Computational Science Research Center, Beijing 100084, China}
\author{Jiangfeng Du}
\affiliation{Hefei National Laboratory for Physics Sciences at Microscale and Department of
Modern Physics, University of Science and Technology of China, Hefei, Anhui
230026, China}
\author{Wen Yang}
\affiliation{Beijing Computational Science Research Center, Beijing 100084, China}
\email{wenyang@csrc.ac.cn}

\begin{abstract}
We develop a quantum theory for a variety of nuclear spin dynamics such as
dephasing, relaxation, squeezing, and narrowing due to the hyperfine
interaction with a generic, dissipative electronic system. The first-order
result of our theory reproduces and generalizes the nonlinear Hamiltonian for
nuclear spin squeezing [M. S. Rudner \textit{et al}., Phys. Rev. Lett. 107,
206806 (2011)]. The second-order result of our theory provides a good
explanation to the experimentally observed $^{13}$C nuclear spin bath
narrowing in diamond nitrogen-vacancy center [E. Togan \textit{et al}., Nature
478, 497 (2011)].

\end{abstract}

\pacs{03.67.Pp, 71.70.Jp, 76.70.Fz, 03.67.Lx}
\maketitle

Diamond nitrogen-vacancy (NV) center is a leading platform for quantum
computation and sensing at the nanoscale
\cite{DuttScience2007,MazeNature2008,DoldeNatPhys2011,ChildressScience2006,ToganNature2010,NeumannNatPhys2010,NeumannScience2010}%
. An important advantage of the NV center is the long electron spin coherence
time \cite{BalasubramanianNatMater2009}, which is ultimately limited by the
noise from the randomly fluctuating $^{13}\mathrm{C}$ nuclei in ultrapure
samples \cite{ZhaoPRB2012}. To protect the NV spin coherence, dynamical
decoupling \cite{UhrigPRL2007,ZhaoNatNano2012} has achieved remarkable success
in prolonging the NV spin coherence time \cite{BarGillNatCommun2013} for an
\emph{ultrashort }duration $(\sim$ $T_{2}^{\ast}$) around the refocusing
point. To achieve \emph{persistent} coherence protection, especially for
\emph{multiple} coupled spins, a promising approach is to suppress the nuclear
spin noise by narrowing the nuclear spin bath distribution. This approach has
been widely explored and successfully demonstrated in semiconductor quantum
dots\cite{GreilichScience2007,XuNature2009,SunPRL2012,LattaNatPhys2009,HogelePRL2012,VinkNatPhys2009,DanonPRL2009,FolettiNatPhys2009,BluhmPRL2010,UrbaszekRMP2013}%
.

Recently the dynamics of nuclear spins in NV centers is attracting increasing
interest. Experimentally, hyperfine induced nuclear spin decoherence and
relaxation
\cite{DuttScience2007,JiangPRL2008,FischerPRB2013,FischerPRL2013,KingPRB2010,LondonPRL2013,BelthangadyPRL2013,DreauPRB2012,JacquesPRL2009,WangNatCommun2013,DreauPRL2013}
have been studied and $^{13}$C nuclear spin bath narrowing has been observed
\cite{ToganNature2011}. Theoretically, despite many works on the nuclear spin
dynamics induced by the \emph{isotropic} contact hyperfine interaction (HFI)
with electrons in quantum dots, most of them are not directly applicable to
the NV center, because the NV spin decoherence is dominated by the\emph{
anisotropic} dipolar HFI with $^{13}$C nuclei. The dipolar HFI does not
conserve the total spin and leads to very different electron-nuclear coupled
dynamics, e.g., the widely used Fermi golden rule approach does not fully
capture the nuclear spin relaxation under quasi-resonant optical pumping when
the HFI is anisotropic \cite{YangPRB2012,YangPRB2013}. Up to now, only the
dynamics of \emph{a few} nuclei strongly coupled to the NV center has been
treated, either by direct numerical modelling
\cite{JiangPRL2008,FischerPRB2013,FischerPRL2013} or by rate equations to
describe the \emph{incoherent} relaxation of the nuclear spin population, with
the rate obtained either phenomenologically
\cite{JacquesPRL2009,ToganNature2011,DreauPRL2013,WangNatCommun2013} or from
the Fermi golden rule \cite{KingPRB2010}. By contrast, narrowing of the
\emph{many} weakly coupled $^{13}$C nuclei, the dominant source of NV spin
decoherence, has not been addressed theoretically. The experimentally observed
narrowing of $^{13}$C nuclei in NV center \cite{ToganNature2011} is consistent
with a theoretical prediction in semiconductor quantum dots
\cite{IsslerPRL2010}, but the specific physical mechanism remains unclear.

In this letter, we develop a quantum theory for the nuclear spin dynamics
induced by general HFI with a dissipative electronic system. This theory has
three distinguishing features compared with previous works. First, instead of
treating only the incoherent nuclear spin relaxation
\cite{YangPRB2012,YangPRB2013}, it include both the diagonal population and
the off-diagonal coherence and can describe a variety of nuclear spin dynamics
such as dephasing \cite{JiangPRL2008}, squeezing \cite{RudnerPRL2011}, and
dynamic polarization and narrowing
\cite{GreilichScience2007,XuNature2009,SunPRL2012,LattaNatPhys2009,HogelePRL2012,VinkNatPhys2009,DanonPRL2009,FolettiNatPhys2009,BluhmPRL2010,UrbaszekRMP2013,IsslerPRL2010}%
. This is highly desirable given the recent advances of electron-nuclei hybrid
quantum registers
\cite{MorleyNatMater2013,TaminiauNatNano2014,WaldherrNature2014}. Second,
instead of treating the entire HFI as a perturbation \cite{DanonPRB2011}, it
treats the longitudinal HFI non-perturbatively, the key to nuclear spin
narrowing \cite{YangPRB2012} and squeezing \cite{RudnerPRL2011}. Third,
without resorting to large electron-nuclear energy mismatch and weak optical
excitation \cite{IsslerPRL2010}, it only assumes the electron-induced nuclear
spin dynamics to be much slower than the electron damping and is applicable to
many electron-nuclear coupled systems, such as single
\cite{BrackerPRL2005,TartakovskiiPRL2007,ChekhovichPRL2010} and double
\cite{KoppensScience2005,BaughPRL2007,PfundPRL2007,ChurchillNatPhys2009,PetersenPRL2013}
quantum dots including quadrupolar interactions
\cite{DzhioevPRL2007,KrebsPRL2010}, as well as NV centers
\cite{JiangPRL2008,ToganNature2011,JacquesPRL2009,FischerPRB2013,FischerPRL2013,KingPRB2010,WangNatCommun2013,ToganNature2011,LondonPRL2013,BelthangadyPRL2013,DreauPRL2013,DreauPRB2012}%
. We exemplify this theory in two paradigmatic examples. The first-order
result reproduces and generalizes the nonlinear Hamiltonian responsible for
nuclear spin squeezing as proposed in Ref. \cite{RudnerPRL2011}. The
second-order result provides a good explanation to the observed $^{13}$C
nuclear spin narrowing \cite{ToganNature2011} in NV center.

We consider many nuclear spins $\{\hat{\mathbf{I}}_{k}\}$ coupled to a
generic, dissipative electron system. The nuclear Hamiltonian $\hat{H}_{N}$
may include the Zeeman term and quadrupolar effect. The electron Hamiltonian
includes multiple energy levels and external control such as optical/microwave
pumping. We \emph{always} work in an appropriate electron rotating frame and
the nuclear spin interaction picture, and decompose the total Hamiltonian
$\hat{H}(t)$ into the time-independent electron part $\hat{H}_{e}$, the
longitudinal HFI $\hat{K}$ that commutes with $\hat{H}_{N}$ and hence induces
no nuclear spin flip between different eigenstates of $\hat{H}_{N}$, and the
transverse HFI $\hat{V}(t)$ that flips the nuclear spins. The coupled system
obeys $\dot{\rho}(t)=-i[\hat{H}_{e}+\hat{K}+\hat{V}(t),\hat{\rho
}(t)]+\mathcal{L}_{e}\hat{\rho}(t)$, with $\mathcal{L}_{e}\hat{\rho}\equiv
\sum_{fi}\gamma_{fi}\mathcal{D}[|f\rangle\langle i|]\hat{\rho}$ for the
electron damping in the Lindblad form $\mathcal{D}[\hat{L}]\hat{\rho}%
\equiv\hat{L}\hat{\rho}\hat{L}^{\dagger}-\{\hat{L}^{\dagger}\hat{L},\hat{\rho
}\}/2$. Here we focus on the electron-induced nuclear spin dynamics and leave
the direct nuclear spin interactions and the intrinsic nuclear spin damping to
the end of our discussion.

To derive a closed equation of motion for the nuclear spin state $\hat
{p}(t)\equiv\operatorname*{Tr}_{e}\hat{\rho}(t)$, we employ the adiabatic
approximation to eliminate the fast electron motion. We introduce the complete
nuclear spin basis set $|\mathbf{m}\rangle\equiv\otimes_{j}|m_{j}\rangle$ as
the common eigenstates of $\hat{H}_{N}$ and $\hat{K}$ with $\hat{K}%
|\mathbf{m}\rangle=\hat{K}_{\mathbf{m}}|\mathbf{m}\rangle$, where $\hat
{K}_{\mathbf{m}}$ is an electron operator, e.g., $\hat{K}_{\mathbf{m}}=\hat
{S}_{z}h_{\mathbf{m}}$ for the contact HFI $\hat{\mathbf{S}}\cdot\sum_{k}%
a_{k}\hat{\mathbf{I}}_{k}\equiv\hat{\mathbf{S}}\cdot\hat{\mathbf{h}}$, with
$h_{\mathbf{m}}\equiv\langle\mathbf{m}|\hat{h}_{z}|\mathbf{m}\rangle$ being
the nuclear field. The ($\mathbf{m},\mathbf{n}$)th block $\hat{\rho
}^{(\mathbf{m},\mathbf{n})}\equiv\langle\mathbf{m}|\hat{\rho}|\mathbf{n}%
\rangle$ of $\hat{\rho}$ obeys
\[
\dot{\rho}^{(\mathbf{m},\mathbf{n})}=\mathcal{L}_{\mathbf{m},\mathbf{n}}%
\hat{\rho}^{(\mathbf{m},\mathbf{n})}-i\frac{\{\hat{\rho}^{(\mathbf{m}%
,\mathbf{n})},\hat{K}_{\mathbf{m,n}}\}}{2}-i\langle\mathbf{m}|[\hat{V}%
,\hat{\rho}]|\mathbf{n}\rangle,
\]
where $\hat{K}_{\mathbf{m,n}}\equiv\hat{K}_{\mathbf{m}}-\hat{K}_{\mathbf{n}}$
and $\mathcal{L}_{\mathbf{m},\mathbf{n}}(\bullet)\equiv-i[\hat{H}%
_{\mathbf{m},\mathbf{n}},\bullet]+\mathcal{L}_{e}(\bullet)$ with $\hat
{H}_{\mathbf{m},\mathbf{n}}\equiv\hat{H}_{e}+(\hat{K}_{\mathbf{m}}+\hat
{K}_{\mathbf{n}})/2$. Tracing over the electron yields the evolution of
$p^{(\mathbf{m},\mathbf{n})}\equiv\langle\mathbf{m}|\hat{p}|\mathbf{n}\rangle
$:
\[
\dot{p}^{(\mathbf{m},\mathbf{n})}=-i\operatorname*{Tr}\nolimits_{e}%
\frac{\{\hat{\rho}^{(\mathbf{m},\mathbf{n})},\hat{K}_{\mathbf{m,n}}\}}%
{2}-i\operatorname*{Tr}\nolimits_{e}\langle\mathbf{m}|[\hat{V},\hat{\rho
}]|\mathbf{n}\rangle.
\]

The above two equations contain four time scales:\ electron evolution and
damping on the time scale $T_{e}$ as driven by $\mathcal{L}_{\mathbf{m}%
,\mathbf{n}}$, nuclear spin precession on the time scale $T_{\mathrm{coh}}$ in
the electron mean field $\langle\hat{K}_{\mathbf{m,n}}\rangle$ and
$\langle\hat{V}(t)\rangle$, nuclear spin dephasing on the time scale $T_{2}$
due to $\hat{K}_{\mathbf{m},\mathbf{n}}$ fluctuation, and nuclear spin
relaxation on the time scale $T_{1}$ due to $\hat{V}(t)$ fluctuation. Any
dynamics much slower than $T_{e}$ can be adiabatically singled out. For
specificity, we consider $T_{e}\ll T_{\mathrm{coh}},T_{1},T_{2}$ and single
out the full dynamics of $\hat{p}(t)$ on the coarse grained time scale $\Delta
t\gg T_{e}$.

To apply the adiabatic approximation, we identify $\hat{p}(t)$ as the slow
variable and other matrix elements of $\hat{\rho}$ as fast variables. We treat
$\mathcal{L}_{\mathbf{m},\mathbf{n}}$ exactly and regard $\hat{K}%
_{\mathbf{m,n}}$ and $\hat{V}(t)$ as first-order small quantities. Carrying
out the adiabatic approximation to successively higher orders (see Sec. A of
\footnote{See supplementary material for derivation of Eqs. (1)--(5) (Sec. A),
exact treatment of the longitudinal HFI $\hat{K}$ (Sec. B), summary of the NV
Hamiltonian under coherent population trapping (Sec. C), and analytical
expression for the NV steady state $\hat{P}_{\mathbf{m},\mathbf{m}}$ (Sec.
D).}) gives the nuclear spin dynamics order by order $\dot{p}=(\dot{p}%
)_{1}+(\dot{p})_{2}+\cdots$. The first-order dynamics
\begin{equation}
(\dot{p}^{(\mathbf{m},\mathbf{n})})_{1}=-i\langle\hat{K}_{\mathbf{m,n}}%
\rangle_{\mathbf{m},\mathbf{n}}p^{(\mathbf{m},\mathbf{n})}-i\operatorname*{Tr}%
\nolimits_{e}\langle\mathbf{m}|[\hat{V},\hat{\rho}_{0}]|\mathbf{n}%
\rangle,\label{EOM1}%
\end{equation}
describes nuclear spin precession in the electron mean fields, which in turn
depends on the nuclear field via $\langle\bullet\rangle_{\mathbf{m}%
,\mathbf{n}}\equiv\operatorname*{Tr}_{e}(\bullet\hat{P}_{\mathbf{m}%
,\mathbf{n}})$. Here $\hat{\rho}_{0}(t)=\sum_{\mathbf{m},\mathbf{n}%
}|\mathbf{m}\rangle\langle\mathbf{n}|p^{(\mathbf{m},\mathbf{n})}(t)\hat
{P}_{\mathbf{m},\mathbf{n}}$ is the zeroth-order approximation to $\hat{\rho
}(t)$ and $\hat{P}_{\mathbf{m},\mathbf{n}}$ is the electron steady state
determined by $\mathcal{L}_{\mathbf{m},\mathbf{n}}\hat{P}_{\mathbf{m}%
,\mathbf{n}}=0$ and $\operatorname*{Tr}_{e}\hat{P}_{\mathbf{m},\mathbf{n}}=1$.
For $\langle\mathbf{p}|\hat{V}(t)|\mathbf{m}\rangle=\hat{V}^{(\mathbf{p}%
,\mathbf{m})}e^{-i\omega_{\mathbf{p},\mathbf{m}}t}$ \footnote{When
$<\mathbf{p}|\hat{V}(t)|\mathbf{m}>$ oscillates at multiple widely
separated frequencies compared with the nuclear spin relaxation and dephasing
rates, the contributions from different frequency components are additive.},
the second-order adiabaic approximation gives the nuclear spin relaxation
\begin{equation}
(\dot{p}^{(\mathbf{m},\mathbf{m})})_{2}=\sum_{\mathbf{p}}(W_{\mathbf{m}%
\leftarrow\mathbf{p}}p^{(\mathbf{p,p})}-W_{\mathbf{p}\leftarrow\mathbf{m}%
}p^{(\mathbf{m},\mathbf{m})})\label{EOM2_D}%
\end{equation}
by the fluctuation of $\hat{V}(t)$, where the transition rate
\begin{equation}
W_{\mathbf{p}\leftarrow\mathbf{m}}=2\operatorname{Re}\int_{0}^{\infty
}e^{i\omega_{\mathbf{p,m}}t}\operatorname*{Tr}\nolimits_{e}\hat{V}%
^{(\mathbf{m,p})}e^{\mathcal{L}_{\mathbf{p},\mathbf{m}}t}\hat{V}%
^{(\mathbf{p},\mathbf{m})}\hat{P}_{\mathbf{m},\mathbf{m}}dt\label{WPM}%
\end{equation}
is a generalized non-equilibrium fluctuation-dissipation relation and reduces
to Refs. \cite{YangPRB2012,YangPRB2013} when $\hat{V}(t)$ is linear in
$\{\mathbf{\hat{I}}_{n}\}$. For the off-diagonal coherences, we have
\begin{equation}
(\dot{p}^{(\mathbf{m},\mathbf{n})})_{2}=-\left(  \Gamma_{\mathbf{m}%
,\mathbf{n}}^{\varphi}+\frac{1}{2}\sum_{\mathbf{p}}(W_{\mathbf{p}%
\leftarrow\mathbf{n}|\mathbf{m}}+W_{\mathbf{p}\leftarrow\mathbf{m}|\mathbf{n}%
})\right)  p^{(\mathbf{m},\mathbf{n)}},\label{EOM2_ND}%
\end{equation}
where we have neglected a second-order energy correction and electron-mediated
nuclear spin interactions, and
\begin{equation}
\Gamma_{\mathbf{m},\mathbf{n}}^{\varphi}\equiv\operatorname{Re}\int
_{0}^{\infty}\operatorname*{Tr}\nolimits_{e}\tilde{K}_{\mathbf{m},\mathbf{n}%
}e^{\mathcal{L}_{\mathbf{m},\mathbf{n}}t}\tilde{K}_{\mathbf{m},\mathbf{n}}%
\hat{P}_{\mathbf{m},\mathbf{n}}dt\label{GAMMA_PHI}%
\end{equation}
is the pure dephasing induced by the fluctuation of $\tilde{K}_{\mathbf{m}%
,\mathbf{n}}\equiv\hat{K}_{\mathbf{m},\mathbf{n}}-\langle\hat{K}%
_{\mathbf{m},\mathbf{n}}\rangle_{\mathbf{m},\mathbf{n}}$. The expression for
$W_{\mathbf{p}\leftarrow\mathbf{m|n}}$ is slightly involved \cite{Note1}, but
it reduces to $W_{\mathbf{p}\leftarrow\mathbf{m}}$ when the difference between
$\hat{K}_{\mathbf{m}}$ and $\hat{K}_{\mathbf{n}}$ is neglected. In this case
Eqs. (\ref{EOM2_D}) and (\ref{EOM2_ND}) reduce to generalized Lindblad master
equation with nonlinear dependence of nuclear spin precession, dephasing, and
relaxation on the nuclear field. This is the origin of nonlinear nuclear spin
effects such as squeezing and narrowing. The above equations follow from
perturbative treatment of both $\hat{K}_{\mathbf{m,n}}$ and $\hat{V}(t)$ on
the time scale $\Delta t\gg T_{e}$. If we focus on nuclear spin relaxation on
the time scale $\Delta t\gg T_{e},T_{2},T_{\mathrm{coh}}$, then we can treat
$\hat{K}_{\mathbf{m,n}}$ exactly and still derive Eqs. (\ref{EOM2_D}) and
(\ref{WPM}) (see Sec. B of \cite{Note1}), with $\mathcal{L}_{\mathbf{p}%
,\mathbf{m}}$ replaced with $\mathcal{L}_{\mathbf{p},\mathbf{m}}%
^{\mathrm{tot}}\equiv\mathcal{L}_{\mathbf{p},\mathbf{m}}(\cdots)-i\{(\cdots
),\hat{K}_{\mathbf{p},\mathbf{m}}\}/2$ in Eq. (\ref{WPM}).

Now we discuss the nuclear spin transition rate beyond the widely used Fermi
golden rule by evaluating Eq. (\ref{WPM}) analytically via a perturbative
expansion of $\int_{0}^{\infty}e^{i\omega t}e^{\mathcal{L}t}dt=-(\mathcal{L}%
+i\omega)^{-1}\equiv-\mathcal{G}$ (with subscripts $\mathbf{p,m}$ suppressed
for brevity). For this purpose, we divide $\mathcal{L}(\bullet)\equiv
-i[\hat{H},\bullet]+\mathcal{L}_{e}(\bullet)$ into the unperturbed part
$\mathcal{L}^{\mathrm{d}}$ and the perturbation $\mathcal{L}^{\mathrm{nd}}$,
\begin{subequations}
\label{LDND}%
\begin{align}
\mathcal{L}^{\mathrm{d}}(\bullet)  &  \equiv-i[\hat{H}^{\mathrm{d}}%
,\bullet]-\frac{1}{2}\{\hat{\Gamma},\bullet\},\label{LD}\\
\mathcal{L}^{\mathrm{nd}}(\bullet)  &  \equiv-i[\hat{H}^{\mathrm{nd}}%
,\bullet]+\sum_{fi}\gamma_{fi}|f\rangle\langle f|\langle i|\bullet|i\rangle,
\label{LND}%
\end{align}
where $\hat{H}^{\mathrm{d}}=\sum_{i}\varepsilon_{i}|i\rangle\langle i|$
($\hat{H}^{\mathrm{nd}}$) is the diagonal (off-diagonal) part of $\hat{H}$,
the self-energy $-\{\hat{\Gamma},\bullet\}/2$ and the quantum jump $\sum
_{fi}\gamma_{fi}|f\rangle\langle f|\langle i|\bullet|i\rangle$ are the
diagonal and off-diagonal part of $\mathcal{L}_{e}$, respectively, with
$\hat{\Gamma}\equiv\sum_{i}\Gamma_{i}|i\rangle\langle i|$, and $\Gamma
_{i}\equiv\sum_{f}\gamma_{fi}$ the total dephasing rate of $|i\rangle$. For
$||\mathcal{L}^{\mathrm{d}}+i\omega||\gg||\mathcal{L}^{\mathrm{nd}}||$, we use
Dyson equation $\mathcal{G}=\mathcal{G}^{\mathrm{d}}-\mathcal{G}^{\mathrm{d}%
}\mathcal{L}^{\mathrm{nd}}\mathcal{G}$ with $\mathcal{G}^{\mathrm{d}}%
\equiv(\mathcal{L}^{\mathrm{d}}+i\omega)^{-1}$ to obtain%
\end{subequations}
\begin{equation}
W_{\mathbf{p}\leftarrow\mathbf{m}}\approx-2\operatorname{Re}\operatorname*{Tr}%
\nolimits_{e}\hat{V}^{(\mathbf{m,p})}(\mathcal{G}^{\mathrm{d}}-\mathcal{G}%
^{\mathrm{d}}\mathcal{L}^{\mathrm{nd}}\mathcal{G}^{\mathrm{d}})\hat
{V}^{(\mathbf{p},\mathbf{m})}\hat{P}_{\mathbf{m},\mathbf{m}}, \label{WPM2}%
\end{equation}
where $\mathcal{G}^{\mathrm{d}}(\bullet)=i\sum_{j,j^{\prime}}|j^{\prime
}\rangle\langle j^{\prime}|\bullet|j\rangle\langle j|/z_{j^{\prime},j}$ with
$z_{j^{\prime},j}\equiv\varepsilon_{j^{\prime}}-\varepsilon_{j}-\omega
-i(\Gamma_{j^{\prime}}+\Gamma_{j})/2$. As an example, for $\hat{V}%
^{(\mathbf{p},\mathbf{m})}=\lambda|f\rangle\langle i|$ ($f\neq i$),
substituting Eqs. (\ref{LDND}) into Eq. (\ref{WPM2}) gives $W_{\mathbf{p}%
\leftarrow\mathbf{m}}$ as the sum of the Fermi golden rule contribution
$W_{\mathbf{p}\leftarrow\mathbf{m}}^{\mathrm{golden}}=2\pi|\lambda|^{2}\langle
i|\hat{P}_{\mathbf{m},\mathbf{m}}|i\rangle\delta^{((\Gamma_{f}+\Gamma_{i}%
)/2)}(\varepsilon_{f}-\varepsilon_{i}-\omega)$ and the quantum coherent
contribution $W_{\mathbf{p}\leftarrow\mathbf{m}}^{\mathrm{coh}}=2|\lambda
|^{2}\operatorname{Im}\sum_{j}\langle i|\hat{P}_{\mathbf{m},\mathbf{m}%
}|j\rangle\langle j|\hat{H}^{\mathrm{nd}}|i\rangle/(z_{f,i}z_{f,j})$, where
$\delta^{(\gamma)}(\Delta)\equiv(\gamma/\pi)/(\Delta^{2}+\gamma^{2})$ is the
Lorentzian shape function. Typical external control gives rise to nonzero
$\hat{H}^{\mathrm{nd}}$ and $\langle i|\hat{P}_{\mathbf{m},\mathbf{m}%
}|j\rangle$, so $W_{\mathbf{p}\leftarrow\mathbf{m}}^{\mathrm{coh}}$ could be
important and even dominate when $W_{\mathbf{p}\leftarrow\mathbf{m}%
}^{\mathrm{golden}}$ is suppressed.

The above theory is applicable to many situations to describe a variety of
electron-induced nuclear spin dynamics. With the dependence of $\mathcal{L}%
_{\mathbf{m},\mathbf{n}}$ and hence $\hat{P}_{\mathbf{m},\mathbf{n}}$ on
$\hat{K}_{\mathbf{m}}$ and $\hat{K}_{\mathbf{n}}$ neglected, Eqs.
(\ref{EOM1}-\ref{GAMMA_PHI}) describe the independent dynamics of individual
nuclear spins \cite{JiangPRL2008}. Including these dependences allow us to
describe correlated nuclear spin dynamics, such as squeezing
\cite{RudnerPRL2011} by Eq. (\ref{EOM1}) and dynamic polarization and
narrowing \cite{IsslerPRL2010,XuNature2009} by Eq. (\ref{WPM}). Taking as an
example the contact HFI $\hat{\mathbf{S}}\cdot\sum_{k}a_{k}\hat{\mathbf{I}%
}_{k}\equiv\hat{\mathbf{S}}\cdot\hat{\mathbf{h}}$ with an electron under
continuous pumping, we identify $\hat{K}=\hat{S}_{z}\hat{h}_{z}$ and neglect
the fast oscillating $\hat{V}(t)$ term. The first-order dynamics in Eq.
(\ref{EOM1}) gives
\[
(\dot{p}^{(\mathbf{m},\mathbf{n)}})_{1}\approx-i(h_{\mathbf{m}}\langle\hat
{S}_{z}\rangle_{\mathbf{m,m}}-h_{\mathbf{n}}\langle\hat{S}_{z}\rangle
_{\mathbf{n,n}})p^{(\mathbf{m},\mathbf{n})}%
\]
for strong electron damping $1/T_{e}\gg|h_{\mathbf{m}}-h_{\mathbf{n}}|$, where
$h_{\mathbf{m}}\equiv\langle\mathbf{m}|\hat{h}_{z}|\mathbf{m}\rangle$. This is
equivalent to $\dot{p}=-i[\hat{p},\hat{H}_{\mathrm{eff}}]$ driven by the
Hamiltonian $\hat{H}_{\mathrm{eff}}\equiv\hat{h}_{z}\langle\hat{S}_{z}%
\rangle_{\hat{h}_{z}}$ with $\langle\hat{S}_{z}\rangle_{\hat{h}_{z}}%
=\sum_{\mathbf{m}}|\mathbf{m}\rangle\langle\mathbf{m|}\langle\hat{S}%
_{z}\rangle_{\mathbf{m,m}}$. Such electron-induced nonlinear\ nuclear spin
Hamiltonian could lead to nuclear spin squeezing, as pineered in Ref.
\cite{RudnerPRL2011}, with a semi-phenomelogical derivation of $\hat
{H}_{\mathrm{eff}}$ for the electron under ESR. Here our first-order result
provides an alternative, microscopic derivation for general electron pumping.

\begin{figure}[ptb]
\includegraphics[width=0.9\columnwidth]{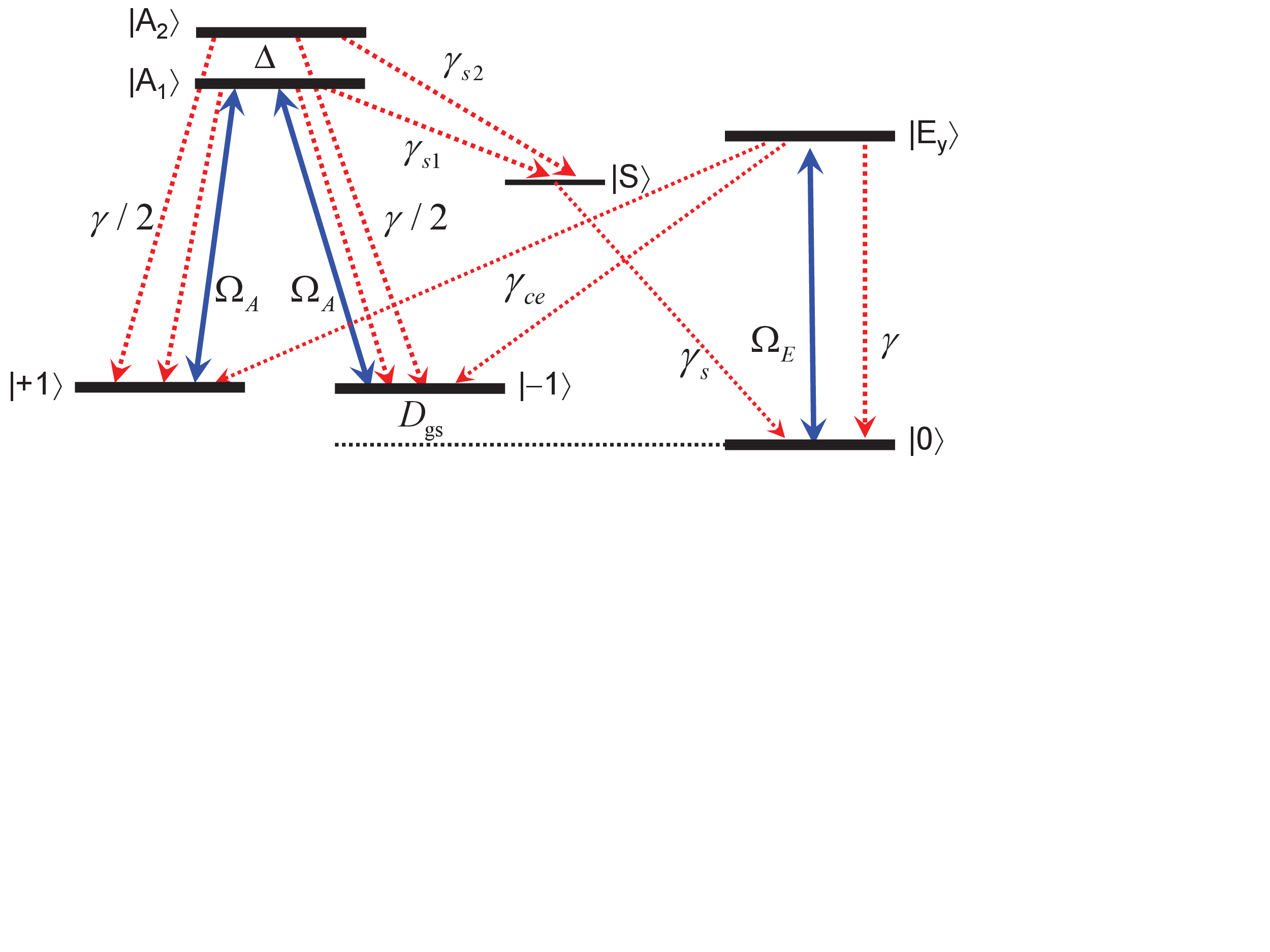}\caption{(color
online) NV center under CPT at low temperature \cite{ToganNature2011}. The
solid (dashed) arrows denote laser excitation (Lindblad damping). The
parameters $\gamma=1/(12\ \mathrm{ns})$, $\gamma_{s1}\sim\gamma$, $\gamma
_{s2}\sim\gamma/120$, and $\gamma_{ce}\sim\gamma/800$ are obtained by fitting
the fluorescence data \cite{ToganNature2011}.}%
\label{G_NV}%
\end{figure}

Finally, we apply the theory to explain the $^{13}$C nuclear spin narrowing
observed in NV center under coherent population trapping (CPT) at low
temperature \cite{ToganNature2011}. The NV states consist of a $\Lambda$
subsystem ($|\pm1\rangle$ and $|A_{1}\rangle$) and a two-level subsystem
($|0\rangle$ and $|E_{y}\rangle$), both under resonant optical pumping (Fig.
\ref{G_NV}). Under two-photon resonance (i.e., when $|\pm1\rangle$ are
degenerate), the bright state $|b\rangle$ of the $\Lambda$ subsystem is pumped
into $|A_{1}\rangle$, which decays into (and is trapped in) the dark state
$|d\rangle$. However, the CPT efficiency is degraded by the off-resonant
optical excitation of $|d\rangle$ into $|A_{2}\rangle$. In the rotating frame
of the two lasers, the NV Hamiltonian $\hat{H}_{e}$ consists of the
ground-state Zeeman splitting $g_{e}\mu_{B}B\hat{S}_{g}^{z}\equiv\omega
_{e}\hat{S}_{g}^{z}$, laser detuning $\Delta|A_{2}\rangle\langle A_{2}|$ for
$|\pm1\rangle\rightarrow|A_{2}\rangle$ excitation, optical pumping
$(\Omega_{A}/\sqrt{2})(|A_{1}\rangle\langle b|+i|A_{2}\rangle\langle
d|+h.c.)+(\Omega_{E}/2)(|E_{y}\rangle\langle0|+h.c.)$, and the strain
term\textbf{ }$\xi_{\perp}(|d\rangle\langle d|-|b\rangle\langle b|)$ (see
\cite{ToganNature2011} or Sec. C of \cite{Note1}). The excited states undergo
spontaneous emission within each subsystem, non-radiative decay between
different subsystems, and pure dephasing $\gamma_{\varphi}$ for each excited
state. Since $\gamma_{s1}\gg\gamma_{ce}$, the population of the excited states
is mostly in $|E_{y}\rangle$.

The NV ground and excited state spins $\hat{\mathbf{S}}_{g}$ and
$\hat{\mathbf{S}}_{e}$ are coupled to the on-site $^{14}$N nucleus
$\hat{\mathbf{I}}_{0}$ via contact HFI $A_{g}\hat{\mathbf{S}}_{g}\cdot
\hat{\mathbf{I}}_{0}+A_{e}\hat{\mathbf{\newline S}}_{e}\cdot\hat{\mathbf{I}%
}_{0}$, where $A_{e}\approx40\ \mathrm{MHz}$ \cite{FuchsPRL2008}, and
$A_{g}\approx2.2\ \mathrm{MHz}$ \cite{DohertyPRB2012}. The total NV spin
$\hat{\mathbf{S}}\equiv\hat{\mathbf{S}}_{g}+\hat{\mathbf{S}}_{e}$ is coupled
to the surrounding $^{13}$C nuclei $\{\hat{\mathbf{I}}_{n}\}$ via dipolar HFI
$\sum_{n=1}^{N}\hat{\mathbf{S}}\cdot\mathbf{A}_{n}\cdot\hat{\mathbf{I}}_{n}$.
For small magnetic field, $\hat{\mathbf{I}}_{0}$ is quantized along the N-V
axis ($z$ axis) by the mean field $A_{g}\langle\hat{\mathbf{S}}_{g}%
\rangle+A_{e}\langle\hat{\mathbf{S}}_{e}\rangle$, which is constant along the
$z$ axis and fast oscillating in the $xy$ plane. Similarly, the $n$th $^{13}$C
nucleus $\hat{\mathbf{I}}_{n}$ is quantized along $\mathbf{e}_{z}%
\cdot\mathbf{A}_{n}$ by the mean field $\langle\hat{\mathbf{S}}\rangle
\cdot\mathbf{A}_{n}$. For convenience, we introduce local Cartesian
coordinates $(\mathbf{e}_{n,x},\mathbf{e}_{n,y},\mathbf{e}_{n,z})$ for the
$n$th $^{13}$C nucleus with $\mathbf{e}_{n,z}=\mathbf{e}_{z}\cdot
\mathbf{A}_{n}/|\mathbf{e}_{z}\cdot\mathbf{A}_{n}|$ and decompose the HFI into
the longitudinal part $\hat{K}\equiv\hat{S}_{g}^{z}\hat{h}_{z}$ and the
transverse part \footnote{As $\hat{S}_{e}^{z}$ and $\mathbf{S}_{\perp}$
exhibit vanishingly small low-frequency fluctuation, the term $\propto\hat
{S}_{e}^{z}$ in $\hat{K}$ and $(\mathbf{S}_{\perp}\cdot\mathbf{A}_{n}%
\cdot\mathbf{e}_{n,z})\hat{I}_{n}^{z}$ in $\hat{V}$ induce negligibly small
nuclear spin dephasing compared with the term $\propto\hat{S}_{g}^{z}$ in
$\hat{K}$ and hence is neglected.}%
\begin{equation}
\hat{V}\equiv(A_{g}\hat{\mathbf{S}}_{g,\perp}+A_{e}\hat{\mathbf{S}}_{e,\perp
})\cdot\hat{\mathbf{I}}_{0,\perp}+\sum_{n=1,2,\cdots,N}\hat{\mathbf{S}}%
_{\perp}\cdot\mathbf{A}_{n}\cdot\hat{\mathbf{I}}_{n,\perp},
\label{V_TRANSVERSE}%
\end{equation}
where $\hat{\mathbf{I}}_{n,\perp}\equiv\hat{I}_{n}^{x}\mathbf{e}_{n,x}+\hat
{I}_{n}^{y}\mathbf{e}_{n,y}$, $\hat{I}_{n}^{\alpha}\equiv\hat{\mathbf{I}}%
_{n}\cdot\mathbf{e}_{n,\alpha}$, and $\hat{h}_{z}=A_{g}\hat{I}_{0}^{z}%
+\sum_{n}|\mathbf{e}_{z}\cdot\mathbf{A}_{n}|\hat{I}_{n}^{z}$ is the nuclear
field. In the rotating frame, the total density matrix $\hat{\rho}(t)$ obeys
$\dot{\rho}(t)=-i[\hat{H}_{e}+\hat{K}+\hat{V}(t),\hat{\rho}(t)]+\mathcal{L}%
_{e}\hat{\rho}$, where $\hat{V}(t)$ is the transverse HFI Eq.
(\ref{V_TRANSVERSE}) transformed into the rotating frame, and $\mathcal{L}%
_{e}$ accounts for NV damping in the Lindblad form.

According to the general theory, we define the nuclear spin basis
$|\mathbf{m}\rangle\equiv\otimes_{n=0}^{N}|m_{n}\rangle$ as the product of
eigenstates of each nucleus: $n=0$ for $^{14}$N (quantized along
$\mathbf{e}_{z}$) and $n=1,2,\cdots,N$ for $^{13}$C (quantized along the local
axis $\mathbf{e}_{n,z}$). Each state $|\mathbf{m}\rangle$ is an eigenstate of
$\hat{K}$, i.e., $\hat{K}|\mathbf{m}\rangle\equiv\hat{S}_{g}^{z}h_{\mathbf{m}%
}|\mathbf{m}\rangle$ with $h_{\mathbf{m}}=\langle\mathbf{m}|\hat{h}%
_{z}|\mathbf{m}\rangle$. The NV steady state $\hat{P}_{\mathbf{m},\mathbf{m}}$
is obtained from $\mathcal{L}_{\mathbf{m},\mathbf{m}}\hat{P}_{\mathbf{m}%
,\mathbf{m}}=0$ and $\operatorname*{Tr}\hat{P}_{\mathbf{m},\mathbf{m}}=1$,
where $\mathcal{L}_{\mathbf{m},\mathbf{m}}(\bullet)\equiv-i[\hat{H}_{e}%
+\hat{S}_{g}^{z}h_{\mathbf{m}},\bullet]+\mathcal{L}_{e}(\bullet)$ and $\hat
{H}_{e}+\hat{S}_{g}^{z}h_{\mathbf{m}}=\hat{H}_{e}|_{\omega_{e}\rightarrow
\delta_{\mathbf{m}}\equiv\omega_{e}+h_{\mathbf{m}}}$, i.e., the nuclear field
$h_{\mathbf{m}}$ changes the two-photon detuning from $2\omega_{e}$ to
$2\delta_{\mathbf{m}}$. Since $\hat{V}^{(\mathbf{m}\pm1_{k},\mathbf{m})}(t)$
oscillates at $\mathrm{GHz}$ frequencies $\gg$ NV damping or laser Rabi
frequencies, the nuclear spin transition rates $W_{\mathbf{m}\pm
1_{k}\leftarrow\mathbf{m}}$ are obtained straightforwardly from the
perturbation formula Eq. (\ref{WPM2}). The transition rate $W_{\mathbf{m}%
+1_{0}\leftarrow\mathbf{m}}=W_{\mathbf{m}-1_{0}\leftarrow\mathbf{m}}$ for
$^{14}$N from $|\mathbf{m}\rangle$ to $|\mathbf{m}\pm1_{0}\rangle\propto
\hat{I}_{0}^{\pm}|\mathbf{m}\rangle$ is dominated by the following
contributions from different NV transitions: $A_{g}^{2}\chi_{g}P_{E_{y}E_{y}}$
from $|0\rangle\rightarrow|\pm1\rangle$ and $A_{e}^{2}(\sum_{f}\chi
_{f})P_{E_{y}E_{y}}$ from $|E_{y}\rangle\rightarrow|f\rangle$, where
$P_{E_{y}E_{y}}\equiv\langle E_{y}|\hat{P}_{\mathbf{m},\mathbf{m}}%
|E_{y}\rangle$ is the population on $|E_{y}\rangle$, $f$ runs over
$A_{1},A_{2},E_{1},E_{2}$ states, $\chi_{g}\equiv(\gamma+2\gamma
_{ce})/D_{\mathrm{gs}}^{2}$, and $\chi_{f}\equiv(1/4)(\Gamma_{f}%
+\gamma_{\varphi})/(\varepsilon_{E_{y}}-\varepsilon_{f})^{2}$, with
$\varepsilon_{f}$ the energy of $|f\rangle$ in the laboratory frame. These
transition rates differ from the phenomenlogical expression $A_{e}%
P_{E_{y}E_{y}}$ in Ref. \cite{ToganNature2011}, which only considers the
$^{14}$N flip by the NV transition $|E_{y}\rangle\rightarrow|A_{1}\rangle$.
Similarly, the transition rate $W_{\mathbf{m}\pm1_{n}\leftarrow\mathbf{m}}$ of
the $n$th $^{13}$C nucleus from $|\mathbf{m}\rangle$ to $|\mathbf{m}\pm
1_{n}\rangle\propto\hat{I}_{n}^{\pm}|\mathbf{m}\rangle$ is dominated by the
following contributions: $\chi_{g}(|A_{n,-,-}|^{2}+|A_{n,+,-}|^{2}%
)P_{E_{y},E_{y}}/8$ from $|0\rangle\rightarrow|\pm1\rangle$, $|A_{n,y,-}%
|^{2}(\chi_{A_{1}}+\chi_{E_{1}})P_{E_{y},E_{y}}/2$ from $|E_{y}\rangle
\rightarrow|A_{1}\rangle,|E_{1}\rangle$, and $|A_{n,x,-}|^{2}(\chi_{A_{2}%
}+\chi_{E_{2}})P_{E_{y},E_{y}}/2$ from $|E_{y}\rangle\rightarrow|A_{2}%
\rangle,|E_{2}\rangle$, where $A_{n,\alpha,\beta}\equiv\mathbf{e}_{n,\alpha
}\cdot\mathbf{A}_{n}\cdot\mathbf{e}_{n,\beta}$. Since $P_{E_{y},E_{y}}$
consists of the dominant CPT term $P_{E_{y},E_{y}}^{(0)}\equiv P_{0}%
\delta_{\mathbf{m}}^{2}/(\delta_{\mathbf{m}}^{2}+\delta_{0}^{2})$ and a small
correction $\sim\chi\equiv(\gamma+\gamma_{\varphi})\Omega_{A}^{2}/(4\eta
_{1}\Delta^{2}(\gamma+\gamma_{s1}))$ from the off-resonant excitation
$|d\rangle\rightarrow|A_{2}\rangle$, all the nuclear spin transition rates
$\propto P_{E_{y},E_{y}}$ are minimized at the two-photon resonance
$\delta_{\mathbf{m}}=0$, where $\eta_{1}=\gamma_{ce}/\gamma_{s1}$ and
$\delta_{0}$ is the intrinsic width of the CPT dip (see Sec. D of
\cite{Note1}). Therefore, although much more involved, the weak-field nuclear
spin dynamics in the NV center is essentially similar to that in quantum dot
under a strong magnetic field \cite{IsslerPRL2010}. The differences are
(i)\ different $^{13}$C nuclei are narrowed about different local axis; (ii)
NV ground and excited states all contribute to the nuclear spin flip and
narrowing; (iii) off-resonant excitation $|d\rangle\rightarrow|A_{2}\rangle$
limits the narrowing efficiency, as suggested in Ref. \cite{ToganNature2011}
and discussed below.

\begin{figure}[ptb]
\includegraphics[width=\columnwidth]{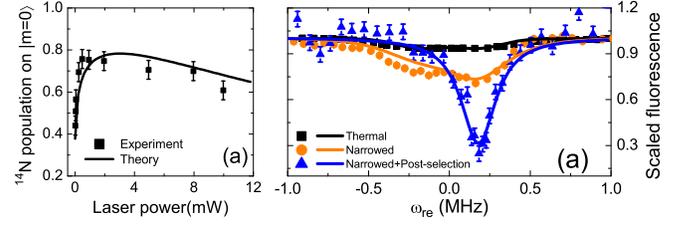} \caption{(color online)
Comparison between experimental \cite{ToganNature2011} (symbols) and
theoretical (lines) results. (a) Population on $|m_{0}=0\rangle$ of $^{14}$N
nucleus. (b) NV fluorescence for different states of the $^{13}$C nuclei. We
choose $\gamma_{C}=2.5\times10^{-2}\ \mathrm{s}^{-1}$, $\gamma_{N}=0,$
$\Omega_{A}=2\ \mathrm{MHz}$, $C=12$, and readout Rabi frequency $\Omega
_{A}^{\mathrm{re}}=3.2$ (black), $10$ (orange), and $8\ \mathrm{MHz}$ (blue).}%
\label{G_COMPARE}%
\end{figure}

The steady state is obtained by solving Eq. (\ref{EOM2_D}), where $\mathbf{n}$
runs over $|\mathbf{m}\pm1_{k}\rangle$ with $k=0,1,\cdots,N$. The nuclear spin
interactions and intrinsic relaxation that are neglected up to now is included
by adding a $^{14}$N depolarization rate $\gamma_{N}$ to $\{W_{\mathbf{m}%
\pm1_{0}\leftarrow\mathbf{m}}\}$ and $^{13}$C depolarization rate $\gamma_{C}$
to $\{W_{\mathbf{m}\pm1_{n}\leftarrow\mathbf{m}}\}$ ($n\geq1$). The calculated
steady state population of $^{14}$N on $|m_{0}=0\rangle$ agrees with the
experiment [Fig. \ref{G_COMPARE}(a)]. At the optimal $\Omega_{A}$
corresponding to maximal population, the calculated $^{14}$N narrowing time
$\sim200$ $\mathrm{\mu s}$ also agrees reasonably with the experimental value
$\sim353\pm34$ $\mathrm{\mu s}$. We further confirm that the decrease of the
population at large $\Omega_{A}$ arises from the off-resonant excitation
$|d\rangle\rightarrow|A_{2}\rangle$, as suggested in Ref.
\cite{ToganNature2011}.

A most important observation is the narrowing of the $^{13}$C nuclei,
manifested as the narrowing of the CPT dip of the NV fluorescence ($\propto$
steady state NV population on $|E_{y}\rangle$) \cite{ToganNature2011}. To
compare with the experiment, we first obtain the nuclear spin steady state
$\hat{p}_{\mathrm{ss}}$ under the experimentally used magnetic field
$\omega_{e}=g_{e}\mu_{B}B=0.18\ \mathrm{MHz}$ and then calculate the $\hat
{p}_{\mathrm{ss}}$-averaged population $\sum_{\mathbf{m}}p_{\mathrm{ss}%
}^{(\mathbf{m},\mathbf{m})}\langle E_{y}|\hat{P}_{\mathbf{m},\mathbf{m}%
}(\omega_{\mathrm{re}})|E_{y}\rangle$ and the post-selected population
$\sum_{\mathbf{m}}p_{\mathrm{ss}}^{(\mathbf{m},\mathbf{m})}e^{-C\langle
E_{y}|\hat{P}_{\mathbf{m},\mathbf{m}}(\omega_{e})|E_{y}\rangle}\langle
E_{y}|\hat{P}_{\mathbf{m},\mathbf{m}}(\omega_{\mathrm{re}})|E_{y}\rangle$
under the readout magnetic field $\omega_{\mathrm{re}}\equiv g_{e}\mu
_{B}B_{\mathrm{re}}$, where $C\langle E_{y}|\hat{P}_{\mathbf{m},\mathbf{m}%
}(\omega_{e})|E_{y}\rangle$ is the average number of collected photons in the
conditioning window. When normalized to unity at large $\omega_{\mathrm{re}}$,
the calculated populations agree with the experimental fluorescence [Fig.
\ref{G_COMPARE}(b)]. To gain a clear understanding of the narrowing, we
neglect $^{14}$N, replace the dipolar HFI tensor $\mathbf{A}_{n}$ by a uniform
one $A_{\parallel}\mathbf{e}_{z}\mathbf{e}_{z}+A_{\perp}(\mathbf{e}%
_{x}\mathbf{e}_{x}+\mathbf{e}_{y}\mathbf{e}_{y})$, set\textbf{ }$\omega_{e}%
=0$, and use Fokker-Planck equation \cite{YangPRB2012,YangPRB2013} to obtain
the distribution $p_{\mathrm{ss}}(h)\equiv\operatorname*{Tr}\delta(\hat{h}%
_{C}-h)\hat{p}_{\mathrm{ss}}$ of the $^{13}$C nuclear field $\hat{h}%
_{C}=A_{\parallel}\sum_{n}\hat{I}_{n}^{z}$:%

\begin{equation}
p_{\mathrm{ss}}(h)\propto\left(  1+\frac{R}{R+\Gamma_{\mathrm{dep}}}%
\frac{\delta_{0}^{2}}{h^{2}+\delta_{s}^{2}}\right)  e^{-h^{2}/(2\sigma
_{\mathrm{eq}}^{2})}, \label{DISTRIBUTION}%
\end{equation}
where $\sigma_{\mathrm{eq}}=\sqrt{N}A_{\parallel}/2$ is the fluctuation of
$\hat{h}_{C}$ in thermal equilibrium, $R=(\sum_{f}\chi_{f}+\chi_{g})A_{\perp
}^{2}P_{0}(1-2\chi)$ is the typical $^{13}$C spin-flip rate, $\Gamma
_{\mathrm{dep}}=\gamma_{C}+\chi R$ is the total $^{13}$C depolarization rate
due to the intrinsic depolarization (rate $\gamma_{C}$) and off-resonant
excitation of $|A_{2}\rangle$, and $\delta_{s}=\sqrt{\Gamma_{\mathrm{dep}%
}/(R+\Gamma_{\mathrm{dep}})}\delta_{0}$. Since $R\gg\Gamma_{\mathrm{dep}}$
under typical experimental conditions, the Lorentzian factor $1/(h^{2}%
+\delta_{s}^{2})$ creates a sharp peak in $p_{\mathrm{ss}}(h)$ around $h=0$
with a typical width $\delta_{s}\ll\delta_{0}$. This makes the steady-state
fluctuation $\sigma=(\langle\hat{h}_{C}^{2}\rangle-\langle\hat{h}_{C}%
\rangle^{2})^{1/2}$ of $\hat{h}_{C}$ with respect to $p_{\mathrm{ss}}(h)$ much
smaller than $\sigma_{\mathrm{eq}}$, corresponding to $^{13}$C spin bath
narrowing. Equation (\ref{DISTRIBUTION}) also suggests that the narrowing
would be degraded when $\Omega_{A}$ exceeds an optimal value due to the
increase of $\delta_{0}$ and hence $\delta_{s}$ (by power broadening) and
$\Gamma_{\mathrm{dep}}$ (by off-resonant excitation $|d\rangle\rightarrow
|A_{2}\rangle$). Without the strain and for small depolarization $\gamma_{C}$,
we can obtain the optimal narrowing analytically as $(\sigma/\sigma
_{\mathrm{eq}})_{\min}\approx(4\eta_{3}/(\pi\eta_{1}^{2}))^{1/4}\sqrt
{\sigma_{\mathrm{eq}}/\Delta}$, which is achieved at $\delta_{0}/(\sqrt
{2}\sigma_{\mathrm{eq}})=(P_{0}\gamma_{C}/(2R\eta_{3}))^{1/4}(\Delta\eta
_{1}/\sigma_{\mathrm{eq}})^{1/2}$ with $\eta_{3}=\Gamma_{A_{1}}\Gamma_{A_{2}%
}/(\gamma+\gamma_{s1})^{2}$. We find numerically that the strain has very
small influence on the optimal narrowing, although it has some affect at low
pump power.

The authors thank Nan Zhao and L. J. Sham for helpful discussions. This work
was supported by NSFC (Grant No. 11274036 and No. 11322542) and the MOST
(Grant No. 2014CB848700).

\bibliographystyle{apsrev}

\begin{thebibliography}{56}
\expandafter\ifx\csname natexlab\endcsname\relax\def\natexlab#1{#1}\fi
\expandafter\ifx\csname bibnamefont\endcsname\relax
  \def\bibnamefont#1{#1}\fi
\expandafter\ifx\csname bibfnamefont\endcsname\relax
  \def\bibfnamefont#1{#1}\fi
\expandafter\ifx\csname citenamefont\endcsname\relax
  \def\citenamefont#1{#1}\fi
\expandafter\ifx\csname url\endcsname\relax
  \def\url#1{\texttt{#1}}\fi
\expandafter\ifx\csname urlprefix\endcsname\relax\def\urlprefix{URL }\fi
\providecommand{\bibinfo}[2]{#2}
\providecommand{\eprint}[2][]{\url{#2}}

\bibitem[{\citenamefont{Dutt et~al.}(2007)\citenamefont{Dutt, Childress, Jiang,
  Togan, Maze, Jelezko, Zibrov, Hemmer, and Lukin}}]{DuttScience2007}
\bibinfo{author}{\bibfnamefont{M.~V.~G.} \bibnamefont{Dutt}},
  \bibinfo{author}{\bibfnamefont{L.}~\bibnamefont{Childress}},
  \bibinfo{author}{\bibfnamefont{L.}~\bibnamefont{Jiang}},
  \bibinfo{author}{\bibfnamefont{E.}~\bibnamefont{Togan}},
  \bibinfo{author}{\bibfnamefont{J.}~\bibnamefont{Maze}},
  \bibinfo{author}{\bibfnamefont{F.}~\bibnamefont{Jelezko}},
  \bibinfo{author}{\bibfnamefont{A.~S.} \bibnamefont{Zibrov}},
  \bibinfo{author}{\bibfnamefont{P.~R.} \bibnamefont{Hemmer}},
  \bibnamefont{and} \bibinfo{author}{\bibfnamefont{M.~D.} \bibnamefont{Lukin}},
  \bibinfo{journal}{Science} \textbf{\bibinfo{volume}{316}},
  \bibinfo{pages}{1312} (\bibinfo{year}{2007}).

\bibitem[{\citenamefont{Maze et~al.}(2008)\citenamefont{Maze, Stanwix, Hodges,
  Hong, Taylor, Cappellaro, Jiang, Dutt, Togan, Zibrov
  et~al.}}]{MazeNature2008}
\bibinfo{author}{\bibfnamefont{J.~R.} \bibnamefont{Maze}},
  \bibinfo{author}{\bibfnamefont{P.~L.} \bibnamefont{Stanwix}},
  \bibinfo{author}{\bibfnamefont{J.~S.} \bibnamefont{Hodges}},
  \bibinfo{author}{\bibfnamefont{S.}~\bibnamefont{Hong}},
  \bibinfo{author}{\bibfnamefont{J.~M.} \bibnamefont{Taylor}},
  \bibinfo{author}{\bibfnamefont{P.}~\bibnamefont{Cappellaro}},
  \bibinfo{author}{\bibfnamefont{L.}~\bibnamefont{Jiang}},
  \bibinfo{author}{\bibfnamefont{M.~V.~G.} \bibnamefont{Dutt}},
  \bibinfo{author}{\bibfnamefont{E.}~\bibnamefont{Togan}},
  \bibinfo{author}{\bibfnamefont{A.~S.} \bibnamefont{Zibrov}},
  \bibnamefont{et~al.}, \bibinfo{journal}{Nature}
  \textbf{\bibinfo{volume}{455}}, \bibinfo{pages}{644} (\bibinfo{year}{2008}).

\bibitem[{\citenamefont{Dolde et~al.}(2011)\citenamefont{Dolde, Fedder,
  Doherty, Nobauer, Rempp, Balasubramanian, Wolf, Reinhard, Hollenberg, Jelezko
  et~al.}}]{DoldeNatPhys2011}
\bibinfo{author}{\bibfnamefont{F.}~\bibnamefont{Dolde}},
  \bibinfo{author}{\bibfnamefont{H.}~\bibnamefont{Fedder}},
  \bibinfo{author}{\bibfnamefont{M.~W.} \bibnamefont{Doherty}},
  \bibinfo{author}{\bibfnamefont{T.}~\bibnamefont{Nobauer}},
  \bibinfo{author}{\bibfnamefont{F.}~\bibnamefont{Rempp}},
  \bibinfo{author}{\bibfnamefont{G.}~\bibnamefont{Balasubramanian}},
  \bibinfo{author}{\bibfnamefont{T.}~\bibnamefont{Wolf}},
  \bibinfo{author}{\bibfnamefont{F.}~\bibnamefont{Reinhard}},
  \bibinfo{author}{\bibfnamefont{L.~C.~L.} \bibnamefont{Hollenberg}},
  \bibinfo{author}{\bibfnamefont{F.}~\bibnamefont{Jelezko}},
  \bibnamefont{et~al.}, \bibinfo{journal}{Nat. Phys.}
  \textbf{\bibinfo{volume}{7}}, \bibinfo{pages}{459} (\bibinfo{year}{2011}).

\bibitem[{\citenamefont{Childress et~al.}(2006)\citenamefont{Childress,
  Gurudev~Dutt, Taylor, Zibrov, Jelezko, Wrachtrup, Hemmer, and
  Lukin}}]{ChildressScience2006}
\bibinfo{author}{\bibfnamefont{L.}~\bibnamefont{Childress}},
  \bibinfo{author}{\bibfnamefont{M.~V.} \bibnamefont{Gurudev~Dutt}},
  \bibinfo{author}{\bibfnamefont{J.~M.} \bibnamefont{Taylor}},
  \bibinfo{author}{\bibfnamefont{A.~S.} \bibnamefont{Zibrov}},
  \bibinfo{author}{\bibfnamefont{F.}~\bibnamefont{Jelezko}},
  \bibinfo{author}{\bibfnamefont{J.}~\bibnamefont{Wrachtrup}},
  \bibinfo{author}{\bibfnamefont{P.~R.} \bibnamefont{Hemmer}},
  \bibnamefont{and} \bibinfo{author}{\bibfnamefont{M.~D.} \bibnamefont{Lukin}},
  \bibinfo{journal}{Science} \textbf{\bibinfo{volume}{314}},
  \bibinfo{pages}{281} (\bibinfo{year}{2006}).

\bibitem[{\citenamefont{Togan et~al.}(2010)\citenamefont{Togan, Chu, Trifonov,
  Jiang, Maze, Childress, Dutt, Sorensen, Hemmer, Zibrov
  et~al.}}]{ToganNature2010}
\bibinfo{author}{\bibfnamefont{E.}~\bibnamefont{Togan}},
  \bibinfo{author}{\bibfnamefont{Y.}~\bibnamefont{Chu}},
  \bibinfo{author}{\bibfnamefont{A.~S.} \bibnamefont{Trifonov}},
  \bibinfo{author}{\bibfnamefont{L.}~\bibnamefont{Jiang}},
  \bibinfo{author}{\bibfnamefont{J.}~\bibnamefont{Maze}},
  \bibinfo{author}{\bibfnamefont{L.}~\bibnamefont{Childress}},
  \bibinfo{author}{\bibfnamefont{M.~V.~G.} \bibnamefont{Dutt}},
  \bibinfo{author}{\bibfnamefont{A.~S.} \bibnamefont{Sorensen}},
  \bibinfo{author}{\bibfnamefont{P.~R.} \bibnamefont{Hemmer}},
  \bibinfo{author}{\bibfnamefont{A.~S.} \bibnamefont{Zibrov}},
  \bibnamefont{et~al.}, \bibinfo{journal}{Nature}
  \textbf{\bibinfo{volume}{466}}, \bibinfo{pages}{730} (\bibinfo{year}{2010}).

\bibitem[{\citenamefont{Neumann
  et~al.}(2010{\natexlab{a}})\citenamefont{Neumann, Kolesov, Naydenov, Beck,
  Rempp, Steiner, Jacques, Balasubramanian, Markham, Twitchen
  et~al.}}]{NeumannNatPhys2010}
\bibinfo{author}{\bibfnamefont{P.}~\bibnamefont{Neumann}},
  \bibinfo{author}{\bibfnamefont{R.}~\bibnamefont{Kolesov}},
  \bibinfo{author}{\bibfnamefont{B.}~\bibnamefont{Naydenov}},
  \bibinfo{author}{\bibfnamefont{J.}~\bibnamefont{Beck}},
  \bibinfo{author}{\bibfnamefont{F.}~\bibnamefont{Rempp}},
  \bibinfo{author}{\bibfnamefont{M.}~\bibnamefont{Steiner}},
  \bibinfo{author}{\bibfnamefont{V.}~\bibnamefont{Jacques}},
  \bibinfo{author}{\bibfnamefont{G.}~\bibnamefont{Balasubramanian}},
  \bibinfo{author}{\bibfnamefont{M.~L.} \bibnamefont{Markham}},
  \bibinfo{author}{\bibfnamefont{D.~J.} \bibnamefont{Twitchen}},
  \bibnamefont{et~al.}, \bibinfo{journal}{Nat. Phys.}
  \textbf{\bibinfo{volume}{6}}, \bibinfo{pages}{249}
  (\bibinfo{year}{2010}{\natexlab{a}}).

\bibitem[{\citenamefont{Neumann
  et~al.}(2010{\natexlab{b}})\citenamefont{Neumann, Beck, Steiner, Rempp,
  Fedder, Hemmer, Wrachtrup, and Jelezko}}]{NeumannScience2010}
\bibinfo{author}{\bibfnamefont{P.}~\bibnamefont{Neumann}},
  \bibinfo{author}{\bibfnamefont{J.}~\bibnamefont{Beck}},
  \bibinfo{author}{\bibfnamefont{M.}~\bibnamefont{Steiner}},
  \bibinfo{author}{\bibfnamefont{F.}~\bibnamefont{Rempp}},
  \bibinfo{author}{\bibfnamefont{H.}~\bibnamefont{Fedder}},
  \bibinfo{author}{\bibfnamefont{P.~R.} \bibnamefont{Hemmer}},
  \bibinfo{author}{\bibfnamefont{J.}~\bibnamefont{Wrachtrup}},
  \bibnamefont{and} \bibinfo{author}{\bibfnamefont{F.}~\bibnamefont{Jelezko}},
  \bibinfo{journal}{Science} \textbf{\bibinfo{volume}{329}},
  \bibinfo{pages}{542} (\bibinfo{year}{2010}{\natexlab{b}}).

\bibitem[{\citenamefont{Balasubramanian
  et~al.}(2009)\citenamefont{Balasubramanian, Neumann, Twitchen, Markham,
  Kolesov, Mizuochi, Isoya, Achard, Beck, Tissler
  et~al.}}]{BalasubramanianNatMater2009}
\bibinfo{author}{\bibfnamefont{G.}~\bibnamefont{Balasubramanian}},
  \bibinfo{author}{\bibfnamefont{P.}~\bibnamefont{Neumann}},
  \bibinfo{author}{\bibfnamefont{D.}~\bibnamefont{Twitchen}},
  \bibinfo{author}{\bibfnamefont{M.}~\bibnamefont{Markham}},
  \bibinfo{author}{\bibfnamefont{R.}~\bibnamefont{Kolesov}},
  \bibinfo{author}{\bibfnamefont{N.}~\bibnamefont{Mizuochi}},
  \bibinfo{author}{\bibfnamefont{J.}~\bibnamefont{Isoya}},
  \bibinfo{author}{\bibfnamefont{J.}~\bibnamefont{Achard}},
  \bibinfo{author}{\bibfnamefont{J.}~\bibnamefont{Beck}},
  \bibinfo{author}{\bibfnamefont{J.}~\bibnamefont{Tissler}},
  \bibnamefont{et~al.}, \bibinfo{journal}{Nat. Mater.}
  \textbf{\bibinfo{volume}{8}}, \bibinfo{pages}{383} (\bibinfo{year}{2009}).

\bibitem[{\citenamefont{Zhao et~al.}(2012{\natexlab{a}})\citenamefont{Zhao, Ho,
  and Liu}}]{ZhaoPRB2012}
\bibinfo{author}{\bibfnamefont{N.}~\bibnamefont{Zhao}},
  \bibinfo{author}{\bibfnamefont{S.-W.} \bibnamefont{Ho}}, \bibnamefont{and}
  \bibinfo{author}{\bibfnamefont{R.-B.} \bibnamefont{Liu}},
  \bibinfo{journal}{Phys. Rev. B} \textbf{\bibinfo{volume}{85}},
  \bibinfo{pages}{115303} (\bibinfo{year}{2012}{\natexlab{a}}).

\bibitem[{\citenamefont{Uhrig}(2007)}]{UhrigPRL2007}
\bibinfo{author}{\bibfnamefont{G.~S.} \bibnamefont{Uhrig}},
  \bibinfo{journal}{Phys. Rev. Lett.} \textbf{\bibinfo{volume}{98}},
  \bibinfo{pages}{100504} (\bibinfo{year}{2007}).

\bibitem[{\citenamefont{Zhao et~al.}(2012{\natexlab{b}})\citenamefont{Zhao,
  Honert, Schmid, Klas, Isoya, Markham, Twitchen, Jelezko, Liu, Fedder
  et~al.}}]{ZhaoNatNano2012}
\bibinfo{author}{\bibfnamefont{N.}~\bibnamefont{Zhao}},
  \bibinfo{author}{\bibfnamefont{J.}~\bibnamefont{Honert}},
  \bibinfo{author}{\bibfnamefont{B.}~\bibnamefont{Schmid}},
  \bibinfo{author}{\bibfnamefont{M.}~\bibnamefont{Klas}},
  \bibinfo{author}{\bibfnamefont{J.}~\bibnamefont{Isoya}},
  \bibinfo{author}{\bibfnamefont{M.}~\bibnamefont{Markham}},
  \bibinfo{author}{\bibfnamefont{D.}~\bibnamefont{Twitchen}},
  \bibinfo{author}{\bibfnamefont{F.}~\bibnamefont{Jelezko}},
  \bibinfo{author}{\bibfnamefont{R.-B.} \bibnamefont{Liu}},
  \bibinfo{author}{\bibfnamefont{H.}~\bibnamefont{Fedder}},
  \bibnamefont{et~al.}, \bibinfo{journal}{Nat Nano}
  \textbf{\bibinfo{volume}{7}}, \bibinfo{pages}{657}
  (\bibinfo{year}{2012}{\natexlab{b}}).

\bibitem[{\citenamefont{Bar-Gill et~al.}(2013)\citenamefont{Bar-Gill, Pham,
  Jarmola, Budker, and Walsworth}}]{BarGillNatCommun2013}
\bibinfo{author}{\bibfnamefont{N.}~\bibnamefont{Bar-Gill}},
  \bibinfo{author}{\bibfnamefont{L.~M.} \bibnamefont{Pham}},
  \bibinfo{author}{\bibfnamefont{A.}~\bibnamefont{Jarmola}},
  \bibinfo{author}{\bibfnamefont{D.}~\bibnamefont{Budker}}, \bibnamefont{and}
  \bibinfo{author}{\bibfnamefont{R.~L.} \bibnamefont{Walsworth}},
  \bibinfo{journal}{Nat. Commun.} \textbf{\bibinfo{volume}{4}},
  \bibinfo{pages}{1743} (\bibinfo{year}{2013}).

\bibitem[{\citenamefont{Greilich et~al.}(2007)\citenamefont{Greilich, Shabaev,
  Yakovlev, Efros, Yugova, Reuter, Wieck, and Bayer}}]{GreilichScience2007}
\bibinfo{author}{\bibfnamefont{A.}~\bibnamefont{Greilich}},
  \bibinfo{author}{\bibfnamefont{A.}~\bibnamefont{Shabaev}},
  \bibinfo{author}{\bibfnamefont{D.~R.} \bibnamefont{Yakovlev}},
  \bibinfo{author}{\bibfnamefont{A.~L.} \bibnamefont{Efros}},
  \bibinfo{author}{\bibfnamefont{I.~A.} \bibnamefont{Yugova}},
  \bibinfo{author}{\bibfnamefont{D.}~\bibnamefont{Reuter}},
  \bibinfo{author}{\bibfnamefont{A.~D.} \bibnamefont{Wieck}}, \bibnamefont{and}
  \bibinfo{author}{\bibfnamefont{M.}~\bibnamefont{Bayer}},
  \bibinfo{journal}{Science} \textbf{\bibinfo{volume}{317}},
  \bibinfo{pages}{1896} (\bibinfo{year}{2007}).

\bibitem[{\citenamefont{Xu et~al.}(2009)\citenamefont{Xu, Yao, Sun, Steel,
  Bracker, Gammon, and Sham}}]{XuNature2009}
\bibinfo{author}{\bibfnamefont{X.}~\bibnamefont{Xu}},
  \bibinfo{author}{\bibfnamefont{W.}~\bibnamefont{Yao}},
  \bibinfo{author}{\bibfnamefont{B.}~\bibnamefont{Sun}},
  \bibinfo{author}{\bibfnamefont{D.~G.} \bibnamefont{Steel}},
  \bibinfo{author}{\bibfnamefont{A.~S.} \bibnamefont{Bracker}},
  \bibinfo{author}{\bibfnamefont{D.}~\bibnamefont{Gammon}}, \bibnamefont{and}
  \bibinfo{author}{\bibfnamefont{L.~J.} \bibnamefont{Sham}},
  \bibinfo{journal}{Nature} \textbf{\bibinfo{volume}{459}},
  \bibinfo{pages}{1105} (\bibinfo{year}{2009}).

\bibitem[{\citenamefont{Sun et~al.}(2012)\citenamefont{Sun, Chow, Steel,
  Bracker, Gammon, and Sham}}]{SunPRL2012}
\bibinfo{author}{\bibfnamefont{B.}~\bibnamefont{Sun}},
  \bibinfo{author}{\bibfnamefont{C.~M.~E.} \bibnamefont{Chow}},
  \bibinfo{author}{\bibfnamefont{D.~G.} \bibnamefont{Steel}},
  \bibinfo{author}{\bibfnamefont{A.~S.} \bibnamefont{Bracker}},
  \bibinfo{author}{\bibfnamefont{D.}~\bibnamefont{Gammon}}, \bibnamefont{and}
  \bibinfo{author}{\bibfnamefont{L.~J.} \bibnamefont{Sham}},
  \bibinfo{journal}{Phys. Rev. Lett.} \textbf{\bibinfo{volume}{108}},
  \bibinfo{pages}{187401} (\bibinfo{year}{2012}).

\bibitem[{\citenamefont{Latta et~al.}(2009)\citenamefont{Latta, Hogele, Zhao,
  Vamivakas, Maletinsky, Kroner, Dreiser, Carusotto, Badolato, Schuh
  et~al.}}]{LattaNatPhys2009}
\bibinfo{author}{\bibfnamefont{C.}~\bibnamefont{Latta}},
  \bibinfo{author}{\bibfnamefont{A.}~\bibnamefont{Hogele}},
  \bibinfo{author}{\bibfnamefont{Y.}~\bibnamefont{Zhao}},
  \bibinfo{author}{\bibfnamefont{A.~N.} \bibnamefont{Vamivakas}},
  \bibinfo{author}{\bibfnamefont{P.}~\bibnamefont{Maletinsky}},
  \bibinfo{author}{\bibfnamefont{M.}~\bibnamefont{Kroner}},
  \bibinfo{author}{\bibfnamefont{J.}~\bibnamefont{Dreiser}},
  \bibinfo{author}{\bibfnamefont{I.}~\bibnamefont{Carusotto}},
  \bibinfo{author}{\bibfnamefont{A.}~\bibnamefont{Badolato}},
  \bibinfo{author}{\bibfnamefont{D.}~\bibnamefont{Schuh}},
  \bibnamefont{et~al.}, \bibinfo{journal}{Nat. Phys.}
  \textbf{\bibinfo{volume}{5}}, \bibinfo{pages}{758} (\bibinfo{year}{2009}).

\bibitem[{\citenamefont{H\"ogele et~al.}(2012)\citenamefont{H\"ogele, Kroner,
  Latta, Claassen, Carusotto, Bulutay, and Imamoglu}}]{HogelePRL2012}
\bibinfo{author}{\bibfnamefont{A.}~\bibnamefont{H\"ogele}},
  \bibinfo{author}{\bibfnamefont{M.}~\bibnamefont{Kroner}},
  \bibinfo{author}{\bibfnamefont{C.}~\bibnamefont{Latta}},
  \bibinfo{author}{\bibfnamefont{M.}~\bibnamefont{Claassen}},
  \bibinfo{author}{\bibfnamefont{I.}~\bibnamefont{Carusotto}},
  \bibinfo{author}{\bibfnamefont{C.}~\bibnamefont{Bulutay}}, \bibnamefont{and}
  \bibinfo{author}{\bibfnamefont{A.}~\bibnamefont{Imamoglu}},
  \bibinfo{journal}{Phys. Rev. Lett.} \textbf{\bibinfo{volume}{108}},
  \bibinfo{pages}{197403} (\bibinfo{year}{2012}).

\bibitem[{\citenamefont{Vink et~al.}(2009)\citenamefont{Vink, Nowack, Koppens,
  Danon, Nazarov, and Vandersypen}}]{VinkNatPhys2009}
\bibinfo{author}{\bibfnamefont{I.~T.} \bibnamefont{Vink}},
  \bibinfo{author}{\bibfnamefont{K.~C.} \bibnamefont{Nowack}},
  \bibinfo{author}{\bibfnamefont{F.~H.~L.} \bibnamefont{Koppens}},
  \bibinfo{author}{\bibfnamefont{J.}~\bibnamefont{Danon}},
  \bibinfo{author}{\bibfnamefont{Y.~V.} \bibnamefont{Nazarov}},
  \bibnamefont{and} \bibinfo{author}{\bibfnamefont{L.~M.~K.}
  \bibnamefont{Vandersypen}}, \bibinfo{journal}{Nat. Phys.}
  \textbf{\bibinfo{volume}{5}}, \bibinfo{pages}{764} (\bibinfo{year}{2009}).

\bibitem[{\citenamefont{Danon et~al.}(2009)\citenamefont{Danon, Vink, Koppens,
  Nowack, Vandersypen, and Nazarov}}]{DanonPRL2009}
\bibinfo{author}{\bibfnamefont{J.}~\bibnamefont{Danon}},
  \bibinfo{author}{\bibfnamefont{I.~T.} \bibnamefont{Vink}},
  \bibinfo{author}{\bibfnamefont{F.~H.~L.} \bibnamefont{Koppens}},
  \bibinfo{author}{\bibfnamefont{K.~C.} \bibnamefont{Nowack}},
  \bibinfo{author}{\bibfnamefont{L.~M.~K.} \bibnamefont{Vandersypen}},
  \bibnamefont{and} \bibinfo{author}{\bibfnamefont{Y.~V.}
  \bibnamefont{Nazarov}}, \bibinfo{journal}{Phys. Rev. Lett.}
  \textbf{\bibinfo{volume}{103}}, \bibinfo{pages}{046601}
  (\bibinfo{year}{2009}).

\bibitem[{\citenamefont{Foletti et~al.}(2009)\citenamefont{Foletti, Bluhm,
  Mahalu, Umansky, and Yacoby}}]{FolettiNatPhys2009}
\bibinfo{author}{\bibfnamefont{S.}~\bibnamefont{Foletti}},
  \bibinfo{author}{\bibfnamefont{H.}~\bibnamefont{Bluhm}},
  \bibinfo{author}{\bibfnamefont{D.}~\bibnamefont{Mahalu}},
  \bibinfo{author}{\bibfnamefont{V.}~\bibnamefont{Umansky}}, \bibnamefont{and}
  \bibinfo{author}{\bibfnamefont{A.}~\bibnamefont{Yacoby}},
  \bibinfo{journal}{Nat. Phys.} \textbf{\bibinfo{volume}{5}},
  \bibinfo{pages}{903} (\bibinfo{year}{2009}).

\bibitem[{\citenamefont{Bluhm et~al.}(2010)\citenamefont{Bluhm, Foletti,
  Mahalu, Umansky, and Yacoby}}]{BluhmPRL2010}
\bibinfo{author}{\bibfnamefont{H.}~\bibnamefont{Bluhm}},
  \bibinfo{author}{\bibfnamefont{S.}~\bibnamefont{Foletti}},
  \bibinfo{author}{\bibfnamefont{D.}~\bibnamefont{Mahalu}},
  \bibinfo{author}{\bibfnamefont{V.}~\bibnamefont{Umansky}}, \bibnamefont{and}
  \bibinfo{author}{\bibfnamefont{A.}~\bibnamefont{Yacoby}},
  \bibinfo{journal}{Phys. Rev. Lett.} \textbf{\bibinfo{volume}{105}},
  \bibinfo{pages}{216803} (\bibinfo{year}{2010}).

\bibitem[{\citenamefont{Urbaszek et~al.}(2013)\citenamefont{Urbaszek, Marie,
  Amand, Krebs, Voisin, Maletinsky, H\"ogele, and Imamoglu}}]{UrbaszekRMP2013}
\bibinfo{author}{\bibfnamefont{B.}~\bibnamefont{Urbaszek}},
  \bibinfo{author}{\bibfnamefont{X.}~\bibnamefont{Marie}},
  \bibinfo{author}{\bibfnamefont{T.}~\bibnamefont{Amand}},
  \bibinfo{author}{\bibfnamefont{O.}~\bibnamefont{Krebs}},
  \bibinfo{author}{\bibfnamefont{P.}~\bibnamefont{Voisin}},
  \bibinfo{author}{\bibfnamefont{P.}~\bibnamefont{Maletinsky}},
  \bibinfo{author}{\bibfnamefont{A.}~\bibnamefont{H\"ogele}}, \bibnamefont{and}
  \bibinfo{author}{\bibfnamefont{A.}~\bibnamefont{Imamoglu}},
  \bibinfo{journal}{Rev. Mod. Phys.} \textbf{\bibinfo{volume}{85}},
  \bibinfo{pages}{79} (\bibinfo{year}{2013}).

\bibitem[{\citenamefont{Jiang et~al.}(2008)\citenamefont{Jiang, Dutt, Togan,
  Childress, Cappellaro, Taylor, and Lukin}}]{JiangPRL2008}
\bibinfo{author}{\bibfnamefont{L.}~\bibnamefont{Jiang}},
  \bibinfo{author}{\bibfnamefont{M.~V.~G.} \bibnamefont{Dutt}},
  \bibinfo{author}{\bibfnamefont{E.}~\bibnamefont{Togan}},
  \bibinfo{author}{\bibfnamefont{L.}~\bibnamefont{Childress}},
  \bibinfo{author}{\bibfnamefont{P.}~\bibnamefont{Cappellaro}},
  \bibinfo{author}{\bibfnamefont{J.~M.} \bibnamefont{Taylor}},
  \bibnamefont{and} \bibinfo{author}{\bibfnamefont{M.~D.} \bibnamefont{Lukin}},
  \bibinfo{journal}{Phys. Rev. Lett.} \textbf{\bibinfo{volume}{100}},
  \bibinfo{pages}{073001} (\bibinfo{year}{2008}).

\bibitem[{\citenamefont{Fischer
  et~al.}(2013{\natexlab{a}})\citenamefont{Fischer, Jarmola, Kehayias, and
  Budker}}]{FischerPRB2013}
\bibinfo{author}{\bibfnamefont{R.}~\bibnamefont{Fischer}},
  \bibinfo{author}{\bibfnamefont{A.}~\bibnamefont{Jarmola}},
  \bibinfo{author}{\bibfnamefont{P.}~\bibnamefont{Kehayias}}, \bibnamefont{and}
  \bibinfo{author}{\bibfnamefont{D.}~\bibnamefont{Budker}},
  \bibinfo{journal}{Phys. Rev. B} \textbf{\bibinfo{volume}{87}},
  \bibinfo{pages}{125207} (\bibinfo{year}{2013}{\natexlab{a}}).

\bibitem[{\citenamefont{Fischer
  et~al.}(2013{\natexlab{b}})\citenamefont{Fischer, Bretschneider, London,
  Budker, Gershoni, and Frydman}}]{FischerPRL2013}
\bibinfo{author}{\bibfnamefont{R.}~\bibnamefont{Fischer}},
  \bibinfo{author}{\bibfnamefont{C.~O.} \bibnamefont{Bretschneider}},
  \bibinfo{author}{\bibfnamefont{P.}~\bibnamefont{London}},
  \bibinfo{author}{\bibfnamefont{D.}~\bibnamefont{Budker}},
  \bibinfo{author}{\bibfnamefont{D.}~\bibnamefont{Gershoni}}, \bibnamefont{and}
  \bibinfo{author}{\bibfnamefont{L.}~\bibnamefont{Frydman}},
  \bibinfo{journal}{Phys. Rev. Lett.} \textbf{\bibinfo{volume}{111}},
  \bibinfo{pages}{057601} (\bibinfo{year}{2013}{\natexlab{b}}).

\bibitem[{\citenamefont{King et~al.}(2010)\citenamefont{King, Coles, and
  Reimer}}]{KingPRB2010}
\bibinfo{author}{\bibfnamefont{J.~P.} \bibnamefont{King}},
  \bibinfo{author}{\bibfnamefont{P.~J.} \bibnamefont{Coles}}, \bibnamefont{and}
  \bibinfo{author}{\bibfnamefont{J.~A.} \bibnamefont{Reimer}},
  \bibinfo{journal}{Phys. Rev. B} \textbf{\bibinfo{volume}{81}},
  \bibinfo{pages}{073201} (\bibinfo{year}{2010}).

\bibitem[{\citenamefont{London et~al.}(2013)\citenamefont{London, Scheuer, Cai,
  Schwarz, Retzker, Plenio, Katagiri, Teraji, Koizumi, Isoya
  et~al.}}]{LondonPRL2013}
\bibinfo{author}{\bibfnamefont{P.}~\bibnamefont{London}},
  \bibinfo{author}{\bibfnamefont{J.}~\bibnamefont{Scheuer}},
  \bibinfo{author}{\bibfnamefont{J.-M.} \bibnamefont{Cai}},
  \bibinfo{author}{\bibfnamefont{I.}~\bibnamefont{Schwarz}},
  \bibinfo{author}{\bibfnamefont{A.}~\bibnamefont{Retzker}},
  \bibinfo{author}{\bibfnamefont{M.~B.} \bibnamefont{Plenio}},
  \bibinfo{author}{\bibfnamefont{M.}~\bibnamefont{Katagiri}},
  \bibinfo{author}{\bibfnamefont{T.}~\bibnamefont{Teraji}},
  \bibinfo{author}{\bibfnamefont{S.}~\bibnamefont{Koizumi}},
  \bibinfo{author}{\bibfnamefont{J.}~\bibnamefont{Isoya}},
  \bibnamefont{et~al.}, \bibinfo{journal}{Phys. Rev. Lett.}
  \textbf{\bibinfo{volume}{111}}, \bibinfo{pages}{067601}
  (\bibinfo{year}{2013}).

\bibitem[{\citenamefont{Belthangady et~al.}(2013)\citenamefont{Belthangady,
  Bar-Gill, Pham, Arai, Le~Sage, Cappellaro, and
  Walsworth}}]{BelthangadyPRL2013}
\bibinfo{author}{\bibfnamefont{C.}~\bibnamefont{Belthangady}},
  \bibinfo{author}{\bibfnamefont{N.}~\bibnamefont{Bar-Gill}},
  \bibinfo{author}{\bibfnamefont{L.~M.} \bibnamefont{Pham}},
  \bibinfo{author}{\bibfnamefont{K.}~\bibnamefont{Arai}},
  \bibinfo{author}{\bibfnamefont{D.}~\bibnamefont{Le~Sage}},
  \bibinfo{author}{\bibfnamefont{P.}~\bibnamefont{Cappellaro}},
  \bibnamefont{and} \bibinfo{author}{\bibfnamefont{R.~L.}
  \bibnamefont{Walsworth}}, \bibinfo{journal}{Phys. Rev. Lett.}
  \textbf{\bibinfo{volume}{110}}, \bibinfo{pages}{157601}
  (\bibinfo{year}{2013}).

\bibitem[{\citenamefont{Dr\'eau et~al.}(2012)\citenamefont{Dr\'eau, Maze,
  Lesik, Roch, and Jacques}}]{DreauPRB2012}
\bibinfo{author}{\bibfnamefont{A.}~\bibnamefont{Dr\'eau}},
  \bibinfo{author}{\bibfnamefont{J.-R.} \bibnamefont{Maze}},
  \bibinfo{author}{\bibfnamefont{M.}~\bibnamefont{Lesik}},
  \bibinfo{author}{\bibfnamefont{J.-F.} \bibnamefont{Roch}}, \bibnamefont{and}
  \bibinfo{author}{\bibfnamefont{V.}~\bibnamefont{Jacques}},
  \bibinfo{journal}{Phys. Rev. B} \textbf{\bibinfo{volume}{85}},
  \bibinfo{pages}{134107} (\bibinfo{year}{2012}).

\bibitem[{\citenamefont{Jacques et~al.}(2009)\citenamefont{Jacques, Neumann,
  Beck, Markham, Twitchen, Meijer, Kaiser, Balasubramanian, Jelezko, and
  Wrachtrup}}]{JacquesPRL2009}
\bibinfo{author}{\bibfnamefont{V.}~\bibnamefont{Jacques}},
  \bibinfo{author}{\bibfnamefont{P.}~\bibnamefont{Neumann}},
  \bibinfo{author}{\bibfnamefont{J.}~\bibnamefont{Beck}},
  \bibinfo{author}{\bibfnamefont{M.}~\bibnamefont{Markham}},
  \bibinfo{author}{\bibfnamefont{D.}~\bibnamefont{Twitchen}},
  \bibinfo{author}{\bibfnamefont{J.}~\bibnamefont{Meijer}},
  \bibinfo{author}{\bibfnamefont{F.}~\bibnamefont{Kaiser}},
  \bibinfo{author}{\bibfnamefont{G.}~\bibnamefont{Balasubramanian}},
  \bibinfo{author}{\bibfnamefont{F.}~\bibnamefont{Jelezko}}, \bibnamefont{and}
  \bibinfo{author}{\bibfnamefont{J.}~\bibnamefont{Wrachtrup}},
  \bibinfo{journal}{Phys. Rev. Lett.} \textbf{\bibinfo{volume}{102}},
  \bibinfo{pages}{057403} (\bibinfo{year}{2009}).

\bibitem[{\citenamefont{Wang et~al.}(2013)\citenamefont{Wang, Shin, Avalos,
  Seltzer, Budker, Pines, and Bajaj}}]{WangNatCommun2013}
\bibinfo{author}{\bibfnamefont{H.-J.} \bibnamefont{Wang}},
  \bibinfo{author}{\bibfnamefont{C.~S.} \bibnamefont{Shin}},
  \bibinfo{author}{\bibfnamefont{C.~E.} \bibnamefont{Avalos}},
  \bibinfo{author}{\bibfnamefont{S.~J.} \bibnamefont{Seltzer}},
  \bibinfo{author}{\bibfnamefont{D.}~\bibnamefont{Budker}},
  \bibinfo{author}{\bibfnamefont{A.}~\bibnamefont{Pines}}, \bibnamefont{and}
  \bibinfo{author}{\bibfnamefont{V.~S.} \bibnamefont{Bajaj}},
  \bibinfo{journal}{Nat. Commun.} \textbf{\bibinfo{volume}{4}},
  \bibinfo{pages}{1} (\bibinfo{year}{2013}).

\bibitem[{\citenamefont{Dreau et~al.}(2013)\citenamefont{Dreau, Spinicelli,
  Maze, Roch, and Jacques}}]{DreauPRL2013}
\bibinfo{author}{\bibfnamefont{A.}~\bibnamefont{Dreau}},
  \bibinfo{author}{\bibfnamefont{P.}~\bibnamefont{Spinicelli}},
  \bibinfo{author}{\bibfnamefont{J.~R.} \bibnamefont{Maze}},
  \bibinfo{author}{\bibfnamefont{J.-F.} \bibnamefont{Roch}}, \bibnamefont{and}
  \bibinfo{author}{\bibfnamefont{V.}~\bibnamefont{Jacques}},
  \bibinfo{journal}{Phys. Rev. Lett.} \textbf{\bibinfo{volume}{110}},
  \bibinfo{pages}{060502} (\bibinfo{year}{2013}).

\bibitem[{\citenamefont{Togan et~al.}(2011)\citenamefont{Togan, Chu, Imamoglu,
  and Lukin}}]{ToganNature2011}
\bibinfo{author}{\bibfnamefont{E.}~\bibnamefont{Togan}},
  \bibinfo{author}{\bibfnamefont{Y.}~\bibnamefont{Chu}},
  \bibinfo{author}{\bibfnamefont{A.}~\bibnamefont{Imamoglu}}, \bibnamefont{and}
  \bibinfo{author}{\bibfnamefont{M.~D.} \bibnamefont{Lukin}},
  \bibinfo{journal}{Nature} \textbf{\bibinfo{volume}{478}},
  \bibinfo{pages}{497} (\bibinfo{year}{2011}).

\bibitem[{\citenamefont{Yang and Sham}(2012)}]{YangPRB2012}
\bibinfo{author}{\bibfnamefont{W.}~\bibnamefont{Yang}} \bibnamefont{and}
  \bibinfo{author}{\bibfnamefont{L.~J.} \bibnamefont{Sham}},
  \bibinfo{journal}{Phys. Rev. B} \textbf{\bibinfo{volume}{85}},
  \bibinfo{pages}{235319} (\bibinfo{year}{2012}).

\bibitem[{\citenamefont{Yang and Sham}(2013)}]{YangPRB2013}
\bibinfo{author}{\bibfnamefont{W.}~\bibnamefont{Yang}} \bibnamefont{and}
  \bibinfo{author}{\bibfnamefont{L.~J.} \bibnamefont{Sham}},
  \bibinfo{journal}{Phys. Rev. B} \textbf{\bibinfo{volume}{88}},
  \bibinfo{pages}{235304} (\bibinfo{year}{2013}).

\bibitem[{\citenamefont{Issler et~al.}(2010)\citenamefont{Issler, Kessler,
  Giedke, Yelin, Cirac, Lukin, and Imamoglu}}]{IsslerPRL2010}
\bibinfo{author}{\bibfnamefont{M.}~\bibnamefont{Issler}},
  \bibinfo{author}{\bibfnamefont{E.~M.} \bibnamefont{Kessler}},
  \bibinfo{author}{\bibfnamefont{G.}~\bibnamefont{Giedke}},
  \bibinfo{author}{\bibfnamefont{S.}~\bibnamefont{Yelin}},
  \bibinfo{author}{\bibfnamefont{I.}~\bibnamefont{Cirac}},
  \bibinfo{author}{\bibfnamefont{M.~D.} \bibnamefont{Lukin}}, \bibnamefont{and}
  \bibinfo{author}{\bibfnamefont{A.}~\bibnamefont{Imamoglu}},
  \bibinfo{journal}{Phys. Rev. Lett.} \textbf{\bibinfo{volume}{105}},
  \bibinfo{pages}{267202} (\bibinfo{year}{2010}).

\bibitem[{\citenamefont{Rudner et~al.}(2011)\citenamefont{Rudner, Vandersypen,
  Vuleti\ifmmode~\acute{c}\else \'{c}\fi{}, and Levitov}}]{RudnerPRL2011}
\bibinfo{author}{\bibfnamefont{M.~S.} \bibnamefont{Rudner}},
  \bibinfo{author}{\bibfnamefont{L.~M.~K.} \bibnamefont{Vandersypen}},
  \bibinfo{author}{\bibfnamefont{V.}~\bibnamefont{Vuleti\ifmmode~\acute{c}\else
  \'{c}\fi{}}}, \bibnamefont{and} \bibinfo{author}{\bibfnamefont{L.~S.}
  \bibnamefont{Levitov}}, \bibinfo{journal}{Phys. Rev. Lett.}
  \textbf{\bibinfo{volume}{107}}, \bibinfo{pages}{206806}
  (\bibinfo{year}{2011}).

\bibitem[{\citenamefont{Morley et~al.}(2013)\citenamefont{Morley, Lueders,
  Hamed~Mohammady, Balian, Aeppli, Kay, Witzel, Jeschke, and
  Monteiro}}]{MorleyNatMater2013}
\bibinfo{author}{\bibfnamefont{G.~W.} \bibnamefont{Morley}},
  \bibinfo{author}{\bibfnamefont{P.}~\bibnamefont{Lueders}},
  \bibinfo{author}{\bibfnamefont{M.}~\bibnamefont{Hamed~Mohammady}},
  \bibinfo{author}{\bibfnamefont{S.~J.} \bibnamefont{Balian}},
  \bibinfo{author}{\bibfnamefont{G.}~\bibnamefont{Aeppli}},
  \bibinfo{author}{\bibfnamefont{C.~W.~M.} \bibnamefont{Kay}},
  \bibinfo{author}{\bibfnamefont{W.~M.} \bibnamefont{Witzel}},
  \bibinfo{author}{\bibfnamefont{G.}~\bibnamefont{Jeschke}}, \bibnamefont{and}
  \bibinfo{author}{\bibfnamefont{T.~S.} \bibnamefont{Monteiro}},
  \bibinfo{journal}{Nat. Mater.} \textbf{\bibinfo{volume}{12}},
  \bibinfo{pages}{103} (\bibinfo{year}{2013}).

\bibitem[{\citenamefont{Taminiau et~al.}(2014)\citenamefont{Taminiau, Cramer,
  van~der Sar, Dobrovitski, and Hanson}}]{TaminiauNatNano2014}
\bibinfo{author}{\bibfnamefont{T.~H.} \bibnamefont{Taminiau}},
  \bibinfo{author}{\bibfnamefont{J.}~\bibnamefont{Cramer}},
  \bibinfo{author}{\bibfnamefont{T.}~\bibnamefont{van~der Sar}},
  \bibinfo{author}{\bibfnamefont{V.~V.} \bibnamefont{Dobrovitski}},
  \bibnamefont{and} \bibinfo{author}{\bibfnamefont{R.}~\bibnamefont{Hanson}},
  \bibinfo{journal}{Nat Nano} \textbf{\bibinfo{volume}{9}},
  \bibinfo{pages}{171} (\bibinfo{year}{2014}), ISSN \bibinfo{issn}{1748-3387}.

\bibitem[{\citenamefont{Waldherr et~al.}(2014)\citenamefont{Waldherr, Wang,
  Zaiser, Jamali, Schulte-Herbruggen, Abe, Ohshima, Isoya, Du, Neumann
  et~al.}}]{WaldherrNature2014}
\bibinfo{author}{\bibfnamefont{G.}~\bibnamefont{Waldherr}},
  \bibinfo{author}{\bibfnamefont{Y.}~\bibnamefont{Wang}},
  \bibinfo{author}{\bibfnamefont{S.}~\bibnamefont{Zaiser}},
  \bibinfo{author}{\bibfnamefont{M.}~\bibnamefont{Jamali}},
  \bibinfo{author}{\bibfnamefont{T.}~\bibnamefont{Schulte-Herbruggen}},
  \bibinfo{author}{\bibfnamefont{H.}~\bibnamefont{Abe}},
  \bibinfo{author}{\bibfnamefont{T.}~\bibnamefont{Ohshima}},
  \bibinfo{author}{\bibfnamefont{J.}~\bibnamefont{Isoya}},
  \bibinfo{author}{\bibfnamefont{J.~F.} \bibnamefont{Du}},
  \bibinfo{author}{\bibfnamefont{P.}~\bibnamefont{Neumann}},
  \bibnamefont{et~al.}, \bibinfo{journal}{Nature}
  \textbf{\bibinfo{volume}{506}}, \bibinfo{pages}{204} (\bibinfo{year}{2014}),
  ISSN \bibinfo{issn}{0028-0836}.

\bibitem[{\citenamefont{Danon and Nazarov}(2011)}]{DanonPRB2011}
\bibinfo{author}{\bibfnamefont{J.}~\bibnamefont{Danon}} \bibnamefont{and}
  \bibinfo{author}{\bibfnamefont{Y.~V.} \bibnamefont{Nazarov}},
  \bibinfo{journal}{Phys. Rev. B} \textbf{\bibinfo{volume}{83}},
  \bibinfo{pages}{245306} (\bibinfo{year}{2011}).

\bibitem[{\citenamefont{Bracker et~al.}(2005)\citenamefont{Bracker, Stinaff,
  Gammon, Ware, Tischler, Shabaev, Efros, Park, Gershoni, Korenev
  et~al.}}]{BrackerPRL2005}
\bibinfo{author}{\bibfnamefont{A.~S.} \bibnamefont{Bracker}},
  \bibinfo{author}{\bibfnamefont{E.~A.} \bibnamefont{Stinaff}},
  \bibinfo{author}{\bibfnamefont{D.}~\bibnamefont{Gammon}},
  \bibinfo{author}{\bibfnamefont{M.~E.} \bibnamefont{Ware}},
  \bibinfo{author}{\bibfnamefont{J.~G.} \bibnamefont{Tischler}},
  \bibinfo{author}{\bibfnamefont{A.}~\bibnamefont{Shabaev}},
  \bibinfo{author}{\bibfnamefont{A.~L.} \bibnamefont{Efros}},
  \bibinfo{author}{\bibfnamefont{D.}~\bibnamefont{Park}},
  \bibinfo{author}{\bibfnamefont{D.}~\bibnamefont{Gershoni}},
  \bibinfo{author}{\bibfnamefont{V.~L.} \bibnamefont{Korenev}},
  \bibnamefont{et~al.}, \bibinfo{journal}{Phys. Rev. Lett.}
  \textbf{\bibinfo{volume}{94}}, \bibinfo{pages}{047402}
  (\bibinfo{year}{2005}).

\bibitem[{\citenamefont{Tartakovskii et~al.}(2007)\citenamefont{Tartakovskii,
  Wright, Russell, Fal'ko, Van'kov, Skiba-Szymanska, Drouzas, Kolodka,
  Skolnick, Fry et~al.}}]{TartakovskiiPRL2007}
\bibinfo{author}{\bibfnamefont{A.~I.} \bibnamefont{Tartakovskii}},
  \bibinfo{author}{\bibfnamefont{T.}~\bibnamefont{Wright}},
  \bibinfo{author}{\bibfnamefont{A.}~\bibnamefont{Russell}},
  \bibinfo{author}{\bibfnamefont{V.~I.} \bibnamefont{Fal'ko}},
  \bibinfo{author}{\bibfnamefont{A.~B.} \bibnamefont{Van'kov}},
  \bibinfo{author}{\bibfnamefont{J.}~\bibnamefont{Skiba-Szymanska}},
  \bibinfo{author}{\bibfnamefont{I.}~\bibnamefont{Drouzas}},
  \bibinfo{author}{\bibfnamefont{R.~S.} \bibnamefont{Kolodka}},
  \bibinfo{author}{\bibfnamefont{M.~S.} \bibnamefont{Skolnick}},
  \bibinfo{author}{\bibfnamefont{P.~W.} \bibnamefont{Fry}},
  \bibnamefont{et~al.}, \bibinfo{journal}{Phys. Rev. Lett.}
  \textbf{\bibinfo{volume}{98}}, \bibinfo{pages}{026806}
  (\bibinfo{year}{2007}).

\bibitem[{\citenamefont{Chekhovich et~al.}(2010)\citenamefont{Chekhovich,
  Makhonin, Kavokin, Krysa, Skolnick, and Tartakovskii}}]{ChekhovichPRL2010}
\bibinfo{author}{\bibfnamefont{E.~A.} \bibnamefont{Chekhovich}},
  \bibinfo{author}{\bibfnamefont{M.~N.} \bibnamefont{Makhonin}},
  \bibinfo{author}{\bibfnamefont{K.~V.} \bibnamefont{Kavokin}},
  \bibinfo{author}{\bibfnamefont{A.~B.} \bibnamefont{Krysa}},
  \bibinfo{author}{\bibfnamefont{M.~S.} \bibnamefont{Skolnick}},
  \bibnamefont{and} \bibinfo{author}{\bibfnamefont{A.~I.}
  \bibnamefont{Tartakovskii}}, \bibinfo{journal}{Phys. Rev. Lett.}
  \textbf{\bibinfo{volume}{104}}, \bibinfo{pages}{066804}
  (\bibinfo{year}{2010}).

\bibitem[{\citenamefont{Koppens et~al.}(2005)\citenamefont{Koppens, Folk,
  Elzerman, Hanson, van Beveren, Vink, Tranitz, Wegscheider, Kouwenhoven, and
  Vandersypen}}]{KoppensScience2005}
\bibinfo{author}{\bibfnamefont{F.~H.~L.} \bibnamefont{Koppens}},
  \bibinfo{author}{\bibfnamefont{J.~A.} \bibnamefont{Folk}},
  \bibinfo{author}{\bibfnamefont{J.~M.} \bibnamefont{Elzerman}},
  \bibinfo{author}{\bibfnamefont{R.}~\bibnamefont{Hanson}},
  \bibinfo{author}{\bibfnamefont{L.~H.~W.} \bibnamefont{van Beveren}},
  \bibinfo{author}{\bibfnamefont{I.~T.} \bibnamefont{Vink}},
  \bibinfo{author}{\bibfnamefont{H.~P.} \bibnamefont{Tranitz}},
  \bibinfo{author}{\bibfnamefont{W.}~\bibnamefont{Wegscheider}},
  \bibinfo{author}{\bibfnamefont{L.~P.} \bibnamefont{Kouwenhoven}},
  \bibnamefont{and} \bibinfo{author}{\bibfnamefont{L.~M.~K.}
  \bibnamefont{Vandersypen}}, \bibinfo{journal}{Science}
  \textbf{\bibinfo{volume}{309}}, \bibinfo{pages}{1346} (\bibinfo{year}{2005}).

\bibitem[{\citenamefont{Baugh et~al.}(2007)\citenamefont{Baugh, Kitamura, Ono,
  and Tarucha}}]{BaughPRL2007}
\bibinfo{author}{\bibfnamefont{J.}~\bibnamefont{Baugh}},
  \bibinfo{author}{\bibfnamefont{Y.}~\bibnamefont{Kitamura}},
  \bibinfo{author}{\bibfnamefont{K.}~\bibnamefont{Ono}}, \bibnamefont{and}
  \bibinfo{author}{\bibfnamefont{S.}~\bibnamefont{Tarucha}},
  \bibinfo{journal}{Phys. Rev. Lett.} \textbf{\bibinfo{volume}{99}},
  \bibinfo{pages}{096804} (\bibinfo{year}{2007}).

\bibitem[{\citenamefont{Pfund et~al.}(2007)\citenamefont{Pfund, Shorubalko,
  Ensslin, and Leturcq}}]{PfundPRL2007}
\bibinfo{author}{\bibfnamefont{A.}~\bibnamefont{Pfund}},
  \bibinfo{author}{\bibfnamefont{I.}~\bibnamefont{Shorubalko}},
  \bibinfo{author}{\bibfnamefont{K.}~\bibnamefont{Ensslin}}, \bibnamefont{and}
  \bibinfo{author}{\bibfnamefont{R.}~\bibnamefont{Leturcq}},
  \bibinfo{journal}{Phys. Rev. Lett.} \textbf{\bibinfo{volume}{99}},
  \bibinfo{pages}{036801} (\bibinfo{year}{2007}).

\bibitem[{\citenamefont{Churchill et~al.}(2009)\citenamefont{Churchill,
  Bestwick, Harlow, Kuemmeth, Marcos, Stwertka, Watson, and
  Marcus}}]{ChurchillNatPhys2009}
\bibinfo{author}{\bibfnamefont{H.~O.~H.} \bibnamefont{Churchill}},
  \bibinfo{author}{\bibfnamefont{A.~J.} \bibnamefont{Bestwick}},
  \bibinfo{author}{\bibfnamefont{J.~W.} \bibnamefont{Harlow}},
  \bibinfo{author}{\bibfnamefont{F.}~\bibnamefont{Kuemmeth}},
  \bibinfo{author}{\bibfnamefont{D.}~\bibnamefont{Marcos}},
  \bibinfo{author}{\bibfnamefont{C.~H.} \bibnamefont{Stwertka}},
  \bibinfo{author}{\bibfnamefont{S.~K.} \bibnamefont{Watson}},
  \bibnamefont{and} \bibinfo{author}{\bibfnamefont{C.~M.}
  \bibnamefont{Marcus}}, \bibinfo{journal}{Nat Phys}
  \textbf{\bibinfo{volume}{5}}, \bibinfo{pages}{321} (\bibinfo{year}{2009}).

\bibitem[{\citenamefont{Petersen et~al.}(2013)\citenamefont{Petersen, Hoffmann,
  Schuh, Wegscheider, Giedke, and Ludwig}}]{PetersenPRL2013}
\bibinfo{author}{\bibfnamefont{G.}~\bibnamefont{Petersen}},
  \bibinfo{author}{\bibfnamefont{E.~A.} \bibnamefont{Hoffmann}},
  \bibinfo{author}{\bibfnamefont{D.}~\bibnamefont{Schuh}},
  \bibinfo{author}{\bibfnamefont{W.}~\bibnamefont{Wegscheider}},
  \bibinfo{author}{\bibfnamefont{G.}~\bibnamefont{Giedke}}, \bibnamefont{and}
  \bibinfo{author}{\bibfnamefont{S.}~\bibnamefont{Ludwig}},
  \bibinfo{journal}{Phys. Rev. Lett.} \textbf{\bibinfo{volume}{110}},
  \bibinfo{pages}{177602} (\bibinfo{year}{2013}).

\bibitem[{\citenamefont{Dzhioev and Korenev}(2007)}]{DzhioevPRL2007}
\bibinfo{author}{\bibfnamefont{R.~I.} \bibnamefont{Dzhioev}} \bibnamefont{and}
  \bibinfo{author}{\bibfnamefont{V.~L.} \bibnamefont{Korenev}},
  \bibinfo{journal}{Phys. Rev. Lett.} \textbf{\bibinfo{volume}{99}},
  \bibinfo{pages}{037401} (\bibinfo{year}{2007}).

\bibitem[{\citenamefont{Krebs et~al.}(2010)\citenamefont{Krebs, Maletinsky,
  Amand, Urbaszek, Lema\^\i{}tre, Voisin, Marie, and Imamoglu}}]{KrebsPRL2010}
\bibinfo{author}{\bibfnamefont{O.}~\bibnamefont{Krebs}},
  \bibinfo{author}{\bibfnamefont{P.}~\bibnamefont{Maletinsky}},
  \bibinfo{author}{\bibfnamefont{T.}~\bibnamefont{Amand}},
  \bibinfo{author}{\bibfnamefont{B.}~\bibnamefont{Urbaszek}},
  \bibinfo{author}{\bibfnamefont{A.}~\bibnamefont{Lema\^\i{}tre}},
  \bibinfo{author}{\bibfnamefont{P.}~\bibnamefont{Voisin}},
  \bibinfo{author}{\bibfnamefont{X.}~\bibnamefont{Marie}}, \bibnamefont{and}
  \bibinfo{author}{\bibfnamefont{A.}~\bibnamefont{Imamoglu}},
  \bibinfo{journal}{Phys. Rev. Lett.} \textbf{\bibinfo{volume}{104}},
  \bibinfo{pages}{056603} (\bibinfo{year}{2010}).

\bibitem[{Note1()}]{Note1}
Note1, \bibinfo{note}{see supplementary material for derivation of Eqs.
  (1)--(5) (Sec. A), exact treatment of the longitudinal HFI $\protect
  \mathaccentV {hat}05E{K}$ (Sec. B), summary of the NV Hamiltonian under
  coherent population trapping (Sec. C), and analytical expression for the NV
  steady state $\protect \mathaccentV {hat}05E{P}_{\protect \mathbf
  {m},\protect \mathbf {m}}$ (Sec. D).}

\bibitem[{Note2()}]{Note2}
Note2, \bibinfo{note}{when $<\protect \mathbf {p}|\protect \mathaccentV
  {hat}05E{V}(t)|\protect \mathbf {m}>$ oscillates at multiple widely separated
  frequencies compared with the nuclear spin relaxation and dephasing rates,
  the contributions from different frequency components are additive.}

\bibitem[{\citenamefont{Fuchs et~al.}(2008)\citenamefont{Fuchs, Dobrovitski,
  Hanson, Batra, Weis, Schenkel, and Awschalom}}]{FuchsPRL2008}
\bibinfo{author}{\bibfnamefont{G.~D.} \bibnamefont{Fuchs}},
  \bibinfo{author}{\bibfnamefont{V.~V.} \bibnamefont{Dobrovitski}},
  \bibinfo{author}{\bibfnamefont{R.}~\bibnamefont{Hanson}},
  \bibinfo{author}{\bibfnamefont{A.}~\bibnamefont{Batra}},
  \bibinfo{author}{\bibfnamefont{C.~D.} \bibnamefont{Weis}},
  \bibinfo{author}{\bibfnamefont{T.}~\bibnamefont{Schenkel}}, \bibnamefont{and}
  \bibinfo{author}{\bibfnamefont{D.~D.} \bibnamefont{Awschalom}},
  \bibinfo{journal}{Phys. Rev. Lett.} \textbf{\bibinfo{volume}{101}},
  \bibinfo{pages}{117601} (\bibinfo{year}{2008}).

\bibitem[{\citenamefont{Doherty et~al.}(2012)\citenamefont{Doherty, Dolde,
  Fedder, Jelezko, Wrachtrup, Manson, and Hollenberg}}]{DohertyPRB2012}
\bibinfo{author}{\bibfnamefont{M.~W.} \bibnamefont{Doherty}},
  \bibinfo{author}{\bibfnamefont{F.}~\bibnamefont{Dolde}},
  \bibinfo{author}{\bibfnamefont{H.}~\bibnamefont{Fedder}},
  \bibinfo{author}{\bibfnamefont{F.}~\bibnamefont{Jelezko}},
  \bibinfo{author}{\bibfnamefont{J.}~\bibnamefont{Wrachtrup}},
  \bibinfo{author}{\bibfnamefont{N.~B.} \bibnamefont{Manson}},
  \bibnamefont{and} \bibinfo{author}{\bibfnamefont{L.~C.~L.}
  \bibnamefont{Hollenberg}}, \bibinfo{journal}{Phys. Rev. B}
  \textbf{\bibinfo{volume}{85}}, \bibinfo{pages}{205203}
  (\bibinfo{year}{2012}).

\bibitem[{Note3()}]{Note3}
Note3, \bibinfo{note}{as $\protect \mathaccentV {hat}05E{S}_{e}^{z}$ and
  $\protect \mathbf {S}_{\perp }$ exhibit vanishingly small low-frequency
  fluctuation, the term $\propto \protect \mathaccentV {hat}05E{S}_{e}^{z}$ in
  $\protect \mathaccentV {hat}05E{K}$ and $(\protect \mathbf {S}_{\perp }\cdot
  \protect \mathbf {A}_{n}\cdot \protect \mathbf {e}_{n,z})\protect
  \mathaccentV {hat}05E{I}_{n}^{z}$ in $\protect \mathaccentV {hat}05E{V}$
  induce negligibly small nuclear spin dephasing compared with the term
  $\propto \protect \mathaccentV {hat}05E{S}_{g}^{z}$ in $\protect \mathaccentV
  {hat}05E{K}$ and hence is neglected.}

\end{thebibliography}

\end{document}


\title{Supplementary materials for \textquotedblleft Quantum theory of nuclear spin
dynamics in diamond nitrogen-vacancy center\textquotedblright}
\author{Ping Wang}
\affiliation{Hefei National Laboratory for Physics Sciences at Microscale and Department of
Modern Physics, University of Science and Technology of China, Hefei, Anhui
230026, China}
\affiliation{Beijing Computational Science Research Center, Beijing 100084, China}
\author{Jiangfeng Du}
\affiliation{Hefei National Laboratory for Physics Sciences at Microscale and Department of
Modern Physics, University of Science and Technology of China, Hefei, Anhui
230026, China}
\author{Wen Yang}
\affiliation{Beijing Computational Science Research Center, Beijing 100084, China}
\email{wenyang@csrc.ac.cn}
\maketitle

The following sections provide background information related to specific
topics of the main text. Section A provides a detailed derivation of Eqs.
(1)-(5) of the main text. Sec. B provides an exact treatment of the
longitudinal hyperfine interaction (HFI) $\hat{K}$ and derive Eqs. (2) and (3)
of the main text. Sec. C summarizes the NV Hamiltonian under the experimental
condition \cite{ToganNature2011} of coherent population trapping. Sec. D
provides analytical expressions for the steady NV state $\hat{P}%
_{\mathbf{m},\mathbf{m}}$.

\subsection{Derivation of nuclear spin equations of motion}

The starting point is
\begin{align}
\dot{\rho}^{(\mathbf{m},\mathbf{n})}  &  =\mathcal{L}_{\mathbf{m},\mathbf{n}%
}\hat{\rho}^{(\mathbf{m},\mathbf{n})}-i\frac{\{\hat{\rho}^{(\mathbf{m}%
,\mathbf{n})},\hat{K}^{(\mathbf{m},\mathbf{n})}\}}{2}-i\langle\mathbf{m}%
|[\hat{V},\hat{\rho}]|\mathbf{n}\rangle,\label{EOM_RHO}\\
\dot{p}^{(\mathbf{m},\mathbf{n})}  &  =-i\operatorname*{Tr}\nolimits_{e}%
\frac{\{\hat{\rho}^{(\mathbf{m},\mathbf{n})},\hat{K}^{(\mathbf{m},\mathbf{n}%
)}\}}{2}-i\operatorname*{Tr}\nolimits_{e}\langle\mathbf{m}|[\hat{V},\hat{\rho
}]|\mathbf{n}\rangle. \label{EOM_P}%
\end{align}
On the coarse grained time scale $T_{e}\ll\Delta t\ll T_{\mathrm{coh}}%
,T_{1},T_{2}$, the trace $p^{(\mathbf{m},\mathbf{n})}\equiv\operatorname*{Tr}%
_{e}\hat{\rho}^{(\mathbf{m},\mathbf{n})}$ is slowly varying, while other
elements of $\hat{\rho}^{(\mathbf{m},\mathbf{n})}$ almost instantaneously
follows the trace $p^{(\mathbf{m},\mathbf{n})}$. Thus we treat the nuclear
spin density matrix elements $\{p^{(\mathbf{m},\mathbf{n})}\}$ (zeroth-order
quantities) as slow variables and other variables as fast variables and
identify $\hat{K}_{\mathbf{m},\mathbf{n}}$ and $\hat{V}$ as first-order small
quantities. Correspondingly, we decompose $\hat{\rho}=\hat{\rho}_{0}+\hat
{\rho}_{1}+\cdots$, where $\hat{\rho}_{k}$ is a $k$th-order small quantity.
Note that the trace of $\{\hat{\rho}_{0}\}$ is the slow variable $\hat
{p}=\operatorname*{Tr}_{e}\hat{\rho}_{0}$, while other elements of $\hat{\rho
}_{0}$ are fast variables. For consistency, for $k=1,2,\cdots$, we require
$\operatorname*{Tr}\hat{\rho}_{k}=0$, so $\hat{\rho}_{k}$ are fast variables.

Since we always work in the nuclear spin interaction picture, any nuclear spin
evolution is caused by the HFI with the electron. Thus the zeroth-order
dynamics vanishes: $(\dot{p})_{0}=0$.

\subsubsection{First-order dynamics}

The first-order evolution $(\dot{p}^{(\mathbf{m},\mathbf{n}})_{1}$ of the slow
variable $p^{(\mathbf{m},\mathbf{n})}=\langle\mathbf{m}|\hat{p}|\mathbf{n}%
\rangle$ is obtained from Eq. (\ref{EOM_P}) by replacing $\hat{\rho}(t)$ with
$\hat{\rho}_{0}(t)$, i.e., the solution to the zeroth-order version of Eq.
(\ref{EOM_RHO}):%
\[
\dot{\rho}_{0}^{(\mathbf{m},\mathbf{n})}=\mathcal{L}_{\mathbf{m},\mathbf{n}%
}\hat{\rho}_{0}^{(\mathbf{m},\mathbf{n})}.
\]
The trace of this equation gives $\dot{p}^{(\mathbf{m},\mathbf{n})}=0$, i.e.,
we solve for the fast variables contained in $\hat{\rho}_{0}^{(\mathbf{m}%
,\mathbf{n})}$ as functions of the slow variables $p^{(\mathbf{m},\mathbf{n}%
)}$, which are regarded as fixed. Coarse-graining for an interval $T_{e}%
\ll\Delta t\ll T_{\mathrm{coh}},T_{1},T_{2}$ or equivalently calculating the
steady-state solution gives $\hat{\rho}_{0}^{(\mathbf{m},\mathbf{n})}%
(t)=\hat{P}_{\mathbf{m,n}}p^{(\mathbf{m,n})}(t)$, where $\hat{P}%
_{\mathbf{m},\mathbf{n}}=\hat{P}_{\mathbf{n},\mathbf{m}}$ is the
time-independent electron steady state as determined by $\mathcal{L}%
_{\mathbf{m,n}}\hat{P}_{\mathbf{m,n}}=0$ and $\operatorname*{Tr}_{e}\hat
{P}_{\mathbf{m},\mathbf{n}}=1$. By replacing $\hat{\rho}(t)$ with $\hat{\rho
}_{0}(t)=\sum_{\mathbf{m},\mathbf{n}}|\mathbf{m}\rangle\langle\mathbf{n}%
|\hat{\rho}_{0}^{(\mathbf{m},\mathbf{n})}(t)$ in Eq. (\ref{EOM_P}), we obtain
$(\dot{p}^{(\mathbf{m},\mathbf{n}})_{1}$ as given by Eq. (1) of the main text.

\subsubsection{Second-order dynamics}

The second-order evolution $(\dot{p}^{(\mathbf{m},\mathbf{n}})_{2}$ of the
slow variable $p^{(\mathbf{m},\mathbf{n})}$ is obtained from Eq. (\ref{EOM_P})
by replacing $\hat{\rho}(t)$ with $\hat{\rho}_{1}(t)$, the solution to the
first-order version of Eq. (\ref{EOM_RHO}):%
\[
(\dot{\rho}_{0}^{(\mathbf{m},\mathbf{n})})_{1}+\dot{\rho}_{1}^{(\mathbf{m}%
,\mathbf{n})}=\mathcal{L}_{\mathbf{m},\mathbf{n}}\hat{\rho}_{1}^{(\mathbf{m}%
,\mathbf{n})}-i\frac{\{\hat{\rho}_{0}^{(\mathbf{m},\mathbf{n})},\hat
{K}^{(\mathbf{m},\mathbf{n})}\}}{2}-i\langle\mathbf{m}|[\hat{V},\hat{\rho}%
_{0}]\mathbf{n}\rangle,
\]
where $(\dot{\rho}_{0}^{(\mathbf{m},\mathbf{n})})_{1}=P_{\mathbf{m}%
,\mathbf{n}}(\dot{p}^{(\mathbf{m},\mathbf{n})})_{1}$ comes from the
first-order evolution $(\dot{p}^{(\mathbf{m},\mathbf{n})})_{1}$. Substituting
into the above equation and coarse-graining for an interval $1/\gamma_{e}%
\ll\Delta t\ll T_{\mathrm{coh}},T_{1},T_{2}$ (or equivalently calculating the
steady-state solution) gives%
\begin{align*}
\hat{\rho}_{1}^{(\mathbf{m},\mathbf{n})}=ip^{(\mathbf{m},\mathbf{n}%
)}\mathcal{L}_{\mathbf{m},\mathbf{n}}^{-1}\frac{\{P_{\mathbf{m},\mathbf{n}%
},\tilde{K}^{(\mathbf{m},\mathbf{n})}\}}{2}  &  +i\sum_{\mathbf{p}%
}(\mathcal{L}_{\mathbf{m},\mathbf{n}}+i\omega_{\mathbf{m},\mathbf{p}}%
)^{-1}(\hat{V}^{(\mathbf{m},\mathbf{p})}P_{\mathbf{p},\mathbf{n}}-\langle
\hat{V}^{(\mathbf{m},\mathbf{p})}\rangle_{\mathbf{p},\mathbf{n}}%
P_{\mathbf{m},\mathbf{n}})p^{(\mathbf{p},\mathbf{n})}\\
&  -i\sum_{\mathbf{p}}(\mathcal{L}_{\mathbf{m},\mathbf{n}}+i\omega
_{\mathbf{p},\mathbf{n}})^{-1}(P_{\mathbf{m,p}}\hat{V}^{(\mathbf{p}%
,\mathbf{n})}-\langle\hat{V}^{(\mathbf{p},\mathbf{n})}\rangle_{\mathbf{m}%
,\mathbf{p}}P_{\mathbf{m},\mathbf{n}})p^{(\mathbf{m,p})}%
\end{align*}
as a function of the slow variables $\{p^{(\mathbf{m},\mathbf{n})}\}$. Here
$\mathcal{L}_{\mathbf{m},\mathbf{n}}^{-1}\equiv-\int_{0}^{\infty
}e^{\mathcal{L}_{\mathbf{m},\mathbf{n}}t}dt$ and $(\mathcal{L}_{\mathbf{m}%
,\mathbf{n}}+i\omega_{\mathbf{m},\mathbf{p}})^{-1}\equiv-\int_{0}^{\infty
}e^{(\mathcal{L}_{\mathbf{m},\mathbf{n}}+i\omega_{\mathbf{m},\mathbf{p}})t}%
dt$, and we have assumed that $\hat{V}^{(\mathbf{m},\mathbf{n})}(t)=\hat
{V}^{(\mathbf{m},\mathbf{n})}e^{-i\omega_{\mathbf{m},\mathbf{n}}t}$ oscillates
at a single frequency $\omega_{\mathbf{m},\mathbf{n}}$. We can verify the
consistent condition $\operatorname*{Tr}_{e}\hat{\rho}_{1}=0$. Then replacing
$\hat{\rho}(t)$ with $\hat{\rho}_{1}(t)$ in Eq. (\ref{EOM_P}) gives the
desired second-order nuclear spin evolution.

For $\mathbf{m}=\mathbf{n}$, neglecting the coupling of the nuclear spin
population $p^{(\mathbf{m},\mathbf{m})}$ to nuclear spin coherences
$p^{(\mathbf{p},\mathbf{q})}$ $(\mathbf{p}\neq\mathbf{q})$, which amounts to
neglecting the small second-order corrections to the transverse mean field
$\langle\hat{V}(t)\rangle$ and electron-mediated nuclear spin interactions
(they induce nuclear spin diffusion and depolarization, which will be included
phenomenologically at the end of the derivation), we obtain the nuclear spin
relaxation Eq. (2) in the main text. The transition rate from $|\mathbf{m}%
\rangle$ to $|\mathbf{p\rangle}$ induced by $\hat{V}(t)$ is%
\[
W_{\mathbf{p}\leftarrow\mathbf{m}}\equiv-2\operatorname{Re}\operatorname*{Tr}%
\nolimits_{e}\hat{V}^{(\mathbf{m},\mathbf{p})}(\mathcal{L}_{\mathbf{p,m}%
}+i\omega_{\mathbf{p,m}})^{-1}(\hat{V}^{(\mathbf{p},\mathbf{m})}\hat
{P}_{\mathbf{m},\mathbf{m}}-\langle\hat{V}^{(\mathbf{p},\mathbf{m})}%
\rangle_{\mathbf{m},\mathbf{m}}\hat{P}_{\mathbf{p,m}}).
\]
Using $(\mathcal{L}_{\mathbf{p,m}}+i\omega_{\mathbf{p,m}})^{-1}\hat
{P}_{\mathbf{p,m}}=\hat{P}_{\mathbf{p,m}}/(i\omega_{\mathbf{p,m}})$, the
contribution from the second term $\propto\hat{P}_{\mathbf{p,m}}$ is
$(2/\omega_{\mathbf{p,m}})\operatorname{Im}\langle\hat{V}^{(\mathbf{m}%
,\mathbf{p})}\rangle_{\mathbf{p},\mathbf{m}}\langle\hat{V}^{(\mathbf{p}%
,\mathbf{m})}\rangle_{\mathbf{m},\mathbf{m}}$. Since $|\mathbf{p}\rangle$ and
$|\mathbf{m}\rangle$ differs only by a single nuclear spin flip, the
difference between $P_{\mathbf{m},\mathbf{m}}$ and $P_{\mathbf{p},\mathbf{m}}$
are negligible and we recover Eq. (3) in the main text.

For $\mathbf{m}\neq\mathbf{n}$, neglecting the coupling of $p^{(\mathbf{m}%
,\mathbf{n})}$ to other variables, which amounts to neglecting the small
second-order corrections to the transverse mean field and the
electron-mediated nuclear spin interaction, we obtain
\[
(\dot{p}^{(\mathbf{m},\mathbf{n}})_{2}=-i(\delta\epsilon_{\mathbf{m|n}}%
^{(2)}-\delta\epsilon_{\mathbf{n}|\mathbf{m}}^{(2)})p^{(\mathbf{m}%
,\mathbf{n})}-\Gamma_{\mathbf{m},\mathbf{n}}^{\varphi}p^{(\mathbf{m}%
,\mathbf{n})}-\frac{\sum_{\mathbf{p}}(W_{\mathbf{p}\leftarrow\mathbf{n}%
|\mathbf{m}}+W_{\mathbf{p}\leftarrow\mathbf{m}|\mathbf{n}})}{2}%
p^{(\mathbf{m,n})},
\]
where $\Gamma_{\mathbf{m},\mathbf{n}}^{\varphi}$ is given by Eq. (5) of the
main text and%

\begin{align*}
\delta\epsilon_{\mathbf{n}|\mathbf{m}}^{(2)}  &  \equiv-\sum_{\mathbf{p}%
}\operatorname{Im}\operatorname*{Tr}\nolimits_{e}\hat{V}^{(\mathbf{n,p}%
)}(\mathcal{L}_{\mathbf{p},\mathbf{m}}+i\omega_{\mathbf{p,n}})^{-1}(\hat
{V}^{(\mathbf{p,n})}P_{\mathbf{n,m}}-\langle\hat{V}^{(\mathbf{p,n})}%
\rangle_{\mathbf{n},\mathbf{m}}P_{\mathbf{m},\mathbf{p}}),\\
W_{\mathbf{p}\leftarrow\mathbf{m|n}}  &  \equiv-2\operatorname{Re}%
\operatorname*{Tr}\nolimits_{e}\hat{V}^{(\mathbf{m,p})}(\mathcal{L}%
_{\mathbf{p,n}}+i\omega_{\mathbf{p,m}})^{-1}(\hat{V}^{(\mathbf{p,m}%
)}P_{\mathbf{m,n}}-\langle\hat{V}^{(\mathbf{p,m})}\rangle_{\mathbf{m}%
,\mathbf{n}}P_{\mathbf{p,n}}).
\end{align*}

\subsection{Exact treatment of longitudinal hyperfine interaction $\hat{K}$}

For the nuclear spin relaxation time $T_{1}$ much longer than the electron
damping time $T_{e}$, we can adiabatically single out the dynamics of
$\{p^{(\mathbf{m},\mathbf{m})}\}$ on the coarse-grained time scale
$T_{e},T_{\mathrm{coh}},T_{2}\ll\Delta t\ll T_{1}$ by treating both
$\mathcal{L}_{\mathbf{m,n}}$ and $\hat{K}^{(\mathbf{m},\mathbf{n})}$
\emph{exactly} and $\hat{V}$ as a perturbation. The slow variables are
$\{p^{(\mathbf{m},\mathbf{m})}\}$ and others are fast variables. The
zeroth-order steady state solution $\rho_{0}^{(\mathbf{m},\mathbf{n})}=0$
$(\mathbf{m}\neq\mathbf{n})$, so the first-order slow variable dynamics
$(\dot{p}^{(\mathbf{m},\mathbf{m})})_{1}=0$. The first-order steady-state
solution is determined from
\[
\dot{\rho}_{1}^{(\mathbf{m},\mathbf{n})}=\mathcal{L}_{\mathbf{m,n}%
}^{\mathrm{tot}}\hat{\rho}_{1}^{(\mathbf{m},\mathbf{n})}-i\langle
\mathbf{m}|[\hat{V},\hat{\rho}_{0}]|\mathbf{n}\rangle
\]
as%
\[
\hat{\rho}_{1}^{(\mathbf{m},\mathbf{n})}=i\sum_{\mathbf{p}}[(\mathcal{L}%
_{\mathbf{m,n}}^{\mathrm{tot}}+i\omega_{\mathbf{m},\mathbf{p}})^{-1}\hat
{V}^{(\mathbf{m},\mathbf{p})}\hat{\rho}_{0}^{(\mathbf{p},\mathbf{n}%
)}-(\mathcal{L}_{\mathbf{m,n}}^{\mathrm{tot}}+i\omega_{\mathbf{p},\mathbf{n}%
})^{-1}\hat{\rho}_{0}^{(\mathbf{m},\mathbf{p})}\hat{V}^{(\mathbf{p}%
,\mathbf{n})}]
\]
where $\mathcal{L}_{\mathbf{m,n}}^{\mathrm{tot}}(\bullet)\equiv\mathcal{L}%
_{\mathbf{m},\mathbf{n}}(\bullet)-i\{\bullet,\hat{K}^{(\mathbf{m},\mathbf{n}%
)}\}/2$. Replacing $\hat{\rho}(t)$ in Eq. (\ref{EOM_P}) with $\hat{\rho}%
_{1}(t)$ and neglecting the coupling of $p^{(\mathbf{m},\mathbf{m})}$ to
nuclear spin coherences gives Eqs. (2) of the main text, with the transition
rate%
\[
W_{\mathbf{p}\leftarrow\mathbf{m}}=2\operatorname{Re}\int_{0}^{\infty
}e^{i\omega_{\mathbf{p,m}}t}dt\operatorname*{Tr}\nolimits_{e}\hat
{V}^{(\mathbf{m},\mathbf{p})}e^{\mathcal{L}_{\mathbf{p,m}}^{\mathrm{tot}}%
t}\hat{V}^{(\mathbf{p},\mathbf{m})}\hat{P}_{\mathbf{m},\mathbf{m}}%
\]
obtained from Eq. (3) of the main text by replacing $\mathcal{L}%
_{\mathbf{p,m}}^{\mathrm{tot}}$ with $\mathcal{L}_{\mathbf{p,m}}$. In the
perturbation limit, we have $\mathcal{L}_{\mathbf{p,m}}^{\mathrm{tot}}%
\approx\mathcal{L}_{\mathbf{p,m}}$ and recovers Eq. (3) of the main text.

\subsection{NV Hamiltonian under coherent population trapping}

The NV Hamiltonian for the coherent population trapping (CPT) experiment has
been discussed in \cite{ToganNature2011}. Here we reproduce it with greater
detail using our own notations. The NV states of relevance include 3 ground
triplet states $|0\rangle$ (energy $0$), $|\pm1\rangle$ (energy
$D_{\mathrm{gs}}$), 6 excited triplet states $|E_{y}\rangle,|E_{x}\rangle$
(energy $\varepsilon_{E_{x}}=\varepsilon_{E_{y}}$)$,|E_{1}\rangle
,|E_{2}\rangle$ (energy $\varepsilon_{E_{1}}=\varepsilon_{E_{2}}$),
$|A_{1}\rangle$ (energy $\varepsilon_{A_{1}}$)$,|A_{2}\rangle$ (energy
$\varepsilon_{A_{2}}$), and one metastable singlet $|S\rangle$ (energy
$\varepsilon_{S}$). A linearly polarized laser with electric field
$\mathbf{E}_{1}e^{-i\omega_{1}t}/2+c.c.$ and frequency $\omega_{1}%
=\varepsilon_{A_{1}}-D_{\mathrm{gs}}$ resonantly excites the ground states
$|\pm1\rangle$ to the excited state $|A_{1}\rangle$ and, at the same time,
off-resonantly excited $|\pm1\rangle$ to the excited state $|A_{2}\rangle$
with detuning $\Delta=\varepsilon_{A_{2}}-\varepsilon_{A_{1}}$. Another
linearly polarized laser with electric field $\mathbf{E}_{2}e^{-i\omega_{2}%
t}/2+c.c.$ and frequency $\omega_{2}=\varepsilon_{E_{x}}$ resonantly excites
$|0\rangle$ to $|E_{y}\rangle$. The relevant optical transition matrix element
of the electric dipole operator $\mathbf{d}\equiv-e\mathbf{r}$ are $\langle
A_{1}|\mathbf{d}|\pm1\rangle=\pm d_{a,E}\mathbf{e}_{\pm}/(2\sqrt{2})$, $\langle
A_{2}|\mathbf{d}|\pm1\rangle=id_{a,E}\mathbf{e}_{\pm}/(2\sqrt{2})$, and
$\langle E_{y}|\mathbf{d}|0\rangle=d_{a,E}\mathbf{e}_{x}/\sqrt{2}$, where
$\mathbf{e}_{\pm}\equiv\mathbf{e}_{x}\pm i\mathbf{e}_{y}$ and $d_{a,E}$ is the
reduced matrix element of the electric dipole moment
\cite{Dohertyelectronic2011}. Thus the laser coupling Hamiltonian
\[
\hat{H}_{c}(t)=\frac{\Omega_{A}}{2}(e^{i\phi}\hat{\sigma}_{A_{1},+1}%
-e^{-i\phi}\hat{\sigma}_{A_{1},-1}+ie^{i\phi}\hat{\sigma}_{A_{2}%
,+1}+ie^{-i\phi}\hat{\sigma}_{A_{2},-1})e^{-i\omega_{1}t}+\frac{\Omega_{E}}%
{2}e^{-i\omega_{2}t}\hat{\sigma}_{E_{y},0}+h.c.,
\]
where we have defined $\hat{\sigma}_{ij}\equiv|i\rangle\langle j|$,
$\Omega_{E}=\mathbf{E}_{2}\cdot\langle E_{y}|\mathbf{d}|0\rangle
=E_{2,x}d_{a,E}/\sqrt{2}$, and $\Omega_{A}e^{i\phi}\equiv\mathbf{E}_{1}%
\cdot\langle A_{1}|\mathbf{d}|+1\rangle=d_{aE}(E_{1,x}+iE_{1,y})/(2\sqrt{2})$.
Defining the bright state $|b\rangle=(e^{-i\phi}|+1\rangle-e^{i\phi}%
|-1\rangle)/\sqrt{2}$ and dark state $|d\rangle=(e^{-i\phi}|+1\rangle
+e^{i\phi}|-1\rangle)/\sqrt{2}$, the laser coupling Hamiltonian simplifies to
\[
\hat{H}_{c}(t)=\frac{\Omega_{A}}{\sqrt{2}}(\hat{\sigma}_{A_{1},b}+i\hat
{\sigma}_{A_{2},d})e^{-i\omega_{1}t}+\frac{\Omega_{E}}{2}e^{-i\omega_{2}t}%
\hat{\sigma}_{E_{y},0}+h.c..
\]
In the rotating frame $\vert \psi(t)\rangle$ connected to the laboratory frame
$|\psi_{\mathrm{lab}}(t)\rangle$ via $\vert \psi(t)\rangle=e^{i\hat{H}_{0}t}|\psi
_{\mathrm{lab}}(t)\rangle$ with $\hat{H}_{0}=\varepsilon_{E_{x}}(\hat{\sigma
}_{E_{x},E_{x}}+\hat{\sigma}_{E_{y},E_{y}})+\varepsilon_{E_{1}}(\hat{\sigma
}_{E_{1},E_{1}}+\hat{\sigma}_{E_{2},E_{2}})+\varepsilon_{A_{1}}(\hat{\sigma
}_{A_{1},A_{1}}+\hat{\sigma}_{A_{2},A_{2}})+\varepsilon_{S}\hat{\sigma}%
_{S,S}+D_{\mathrm{gs}}(\hat{\sigma}_{1,1}+\hat{\sigma}_{-1,-1})$, the NV
Hamiltonian is%
\[
\Delta\hat{\sigma}_{A_{2},A_{2}}+\frac{\Omega_{A}}{\sqrt{2}}(\hat{\sigma
}_{A_{1},b}+i\hat{\sigma}_{A_{2},d}+h.c.)+\frac{\Omega_{E}}{2}(\hat{\sigma
}_{E_{y},0}+\hat{\sigma}_{0,E_{y}}).
\]
We further include an external magnetic field $B$ along the $z$ (N-V) axis and
uniaxial strain with effective ground strain fields $\xi_{x},\xi_{y},\xi_{z}$.
The former gives rise to ground state Zeeman term $g_{e}\mu_{B}B\hat{S}%
_{g}^{z}\equiv\omega_{e}\hat{S}_{g}^{z}$. The latter induces small corrections
to the energy splitting between $|\pm1\rangle$ and couples $|+1\rangle$ and
$|-1\rangle$ through $\hat{H}_{\mathrm{strain}}\equiv-(\xi_{x}+i\xi_{y}%
)\hat{\sigma}_{1,-1}+h.c=-\xi_{\perp}e^{i\Theta}\hat{\sigma}_{1,-1}+h.c$,
where $\xi_{\perp}\equiv(\xi_{x}^{2}+\xi_{y}^{2})^{1/2}$ and $\Theta$ is
determined by $\xi_{\perp}\cos\Theta=\xi_{x}$ and $\xi_{\perp}\sin\Theta
=\xi_{y}$ (cf. Ref. \cite{ToganNature2011}). The NV Hamiltonian becomes
\[
\hat{H}_{e}=\omega_{e}\hat{S}_{g}^{z}+\hat{H}_{\mathrm{strain}}+\Delta
\hat{\sigma}_{A_{2},A_{2}}+\frac{\Omega_{A}}{\sqrt{2}}(\hat{\sigma}_{A_{1}%
,b}+i\hat{\sigma}_{A_{2},d}+h.c.)+\frac{\Omega_{E}}{2}(\hat{\sigma}_{E_{y}%
,0}+\hat{\sigma}_{0,E_{y}}).
\]
In the basis $|b\rangle,|d\rangle,|0\rangle$, we have $\hat{S}_{g}^{z}%
=\hat{\sigma}_{d,b}+\hat{\sigma}_{b,d}$ and $\hat{\sigma}_{1,-1}=e^{2i\phi
}(\hat{\sigma}_{d,d}-\hat{\sigma}_{b,b}+\hat{\sigma}_{b,d}-\hat{\sigma}%
_{d,b})/2$. Thus the strain term%
\[
\hat{H}_{\mathrm{strain}}=-\xi_{\perp}(\hat{\sigma}_{d,d}-\hat{\sigma}%
_{b,b})\cos(\Theta+2\phi)-i\xi_{\perp}(\hat{\sigma}_{b,d}-\hat{\sigma}%
_{d,b})\sin(\Theta+2\phi)
\]
and the Zeeman term $\omega_{e}\hat{S}_{g}^{z}=\omega_{e}(\hat{\sigma}%
_{d,b}+\hat{\sigma}_{b,d})$ induces coupling between the dark state and the
bright state. The strain-induced coupling is minimized by choosing
$\Theta+2\phi=\pi$ \cite{ToganNature2011}, such that%
\begin{equation}
\hat{H}_{e}=\xi_{\perp}(\hat{\sigma}_{d,d}-\hat{\sigma}_{b,b})+\omega_{e}%
(\hat{\sigma}_{b,d}+\hat{\sigma}_{d,b})+\Delta\hat{\sigma}_{A_{2},A_{2}}%
+\frac{\Omega_{A}}{\sqrt{2}}(\hat{\sigma}_{A_{1},b}+i\hat{\sigma}_{A_{2}%
,d}+h.c.)+\frac{\Omega_{E}}{2}(\hat{\sigma}_{E_{y},0}+\hat{\sigma}_{0,E_{y}}).
\label{HE}%
\end{equation}

\subsection{Analytical expressions for steady NV state}

The NV steady state $\hat{P}_{\mathbf{m},\mathbf{m}}$ (denoted by $\hat{P}$
for brevity) is determined by $\mathcal{L}_{\mathbf{m},\mathbf{m}}\hat{P}=0$
and $\operatorname*{Tr}_{e}\hat{P}=1$, where $\mathcal{L}_{\mathbf{m}%
,\mathbf{m}}(\bullet)\equiv-i[\hat{H}_{e}+\hat{S}_{g}^{z}h_{\mathbf{m}%
},\bullet]+\mathcal{L}_{e}(\bullet)$ and $\hat{H}_{e}+\hat{S}_{g}%
^{z}h_{\mathbf{m}}=\hat{H}_{e}|_{\omega_{e}\rightarrow\delta_{\mathbf{m}%
}\equiv\omega_{e}+h_{\mathbf{m}}}$. Straightforward calculation gives
\begin{align*}
P_{S,S} &  =\frac{2\gamma_{ce}}{\gamma_{s}}P_{E_{y},E_{y}},\\
P_{0,0} &  =\left(  1+\frac{\gamma+2\gamma_{ce}}{W_{E}}\right)  P_{E_{y}%
,E_{y}},\\
\gamma_{s}P_{S,S} &  =\gamma_{s1}P_{A_{1},A_{1}}+\gamma_{s2}P_{A_{2},A_{2}},\\
P_{b,b} &  \approx\eta_{1}\left(  1+\frac{\gamma_{\varphi}}{\Gamma_{_{A_{1}}}%
}+\frac{\gamma+\gamma_{s1}}{W_{A}}\right)  P_{E_{y},E_{y}},
\end{align*}
where $W_{E}\equiv\Omega_{E}^{2}/\Gamma_{A_{1}}$ is the resonant optical
transition rate between $|0\rangle$ and $|E_{y}\rangle$, $W_{A}=\Omega_{A}%
^{2}/\Gamma_{A_{1}}$ is the resonant optical transition rate from
$|\pm1\rangle$ to $|A_{1}\rangle$, and $\eta_{1}\equiv\gamma_{ce}/\gamma_{s1}%
$. We treat the off-resonant $|\pm1\rangle\rightarrow|A_{2}\rangle$ excitation
perturbatively. Up to zeroth order, near the two-photon dark resonance (small
$\delta_{\mathbf{m}}$), we neglect $O(\delta_{\mathbf{m}}^{4})$ and high order
terms and obtain%
\begin{align*}
P_{E_{y},E_{y}}^{(0)} &  =P_{0}\frac{\delta_{\mathbf{m}}^{2}}{\delta
_{\mathbf{m}}^{2}+\delta_{0}^{2}},\\
P_{A_{1},A_{1}}^{(0)} &  =2\eta_{1}P_{E_{y},E_{y}}^{(0)},\\
P_{d,d}^{(0)} &  \approx1-\left(  2+\frac{\gamma+2\gamma_{ce}}{W_{E}}\right)
P_{E_{y},E_{y}}^{(0)},
\end{align*}
where%
\begin{align*}
P_{0} &  \approx\frac{1}{\frac{2}{\eta_{2}}+2\eta_{1}\frac{\gamma+\gamma_{s1}%
}{W_{A}}+\frac{\gamma+2\gamma_{ce}}{W_{E}}},\\
\delta_{0}^{2} &  \approx\frac{\eta_{1}\eta_{2}}{4}\frac{W_{A}^{2}-8W_{A}%
\xi_{\perp}^{2}/\Gamma_{_{A_{1}}}+4\xi^{2}}{\eta_{1}\eta_{2}+\frac{W_{A}%
}{\gamma+\gamma_{s1}}\left(  1+\frac{\eta_{2}}{2}\frac{\gamma+2\gamma_{ce}%
}{W_{E}}\right)  },
\end{align*}
with $\eta_{2}\equiv\gamma_{s}/(\gamma_{s}+\gamma_{ce})$ and we have used
$\xi_{\perp}\ll\Gamma_{_{A_{1}}}$ and $\eta_{1}\ll1$. Note that only
$|E_{y}\rangle,|0\rangle$, and $|d\rangle$ are significantly populated, while
the populations on $|S\rangle,|b\rangle,|A_{1}\rangle$ are much smaller than
that on $|E_{y}\rangle$ since $\gamma_{ce}\ll\gamma_{s},\gamma_{s1},\gamma$.
The leading order correction from the off-resonant excitation to
$|A_{2}\rangle$ are%
\begin{align*}
P_{E_{y}E_{y}}^{(2)} &  \approx\frac{1}{2\eta_{1}}\frac{W_{A_{2}}}%
{\gamma+\gamma_{s1}}P_{dd}^{(0)},\\
P_{A_{1}A_{1}}^{(2)} &  =\frac{W_{A_{2}}}{\gamma+\gamma_{s1}}P_{dd}^{(0)},\\
P_{A_{2}A_{2}}^{(2)} &  =\frac{W_{A_{2}}}{\gamma+\gamma_{s2}}P_{d,d}^{(0)},
\end{align*}
where $W_{A_{2}}=(\Omega_{A}^{2}/2)\Gamma_{_{A_{2}}}/\Delta^{2}$ is the
off-resonant transition rate from $|\pm1\rangle$ to $|A_{2}\rangle$ and we
have used $\gamma_{s2}\ll\gamma,\gamma_{s1}$. Now, with the off-resonant
excitation from $|d\rangle$ to $|A_{2}\rangle$, the populations on the excited
states no longer vanish even on two-photon resonance $\delta_{\mathbf{m}}=0$.
As suggested in Ref. \cite{ToganNature2011}, this off-resonant excitation
limits the efficiency of nuclear spin cooling and narrowing. Under saturated
pumping $W_{E}\gg\gamma$, we have $P_{dd}^{(0)}\approx1-2P_{E_{y},E_{y}}%
^{(0)}$.

\bibliographystyle{apsrev}